\documentclass[twocolumn]{aastex631}

\usepackage{amsmath}
\usepackage{multirow}
\usepackage{algorithm2e}
\usepackage{enumitem}
\usepackage{gensymb}

\begin{document}[]

\title{Spectuner: A Framework for Automated Line Identification of Interstellar Molecules}

\author[0000-0002-7716-1094]{Yisheng Qiu}
\affiliation{Research Center for Astronomical computing, Zhejiang Laboratory, Hangzhou, China}
\email{yishengq@zhejianglab.org}

\author[0000-0002-1466-3484]{Tianwei Zhang}
\affiliation{Research Center for Astronomical computing, Zhejiang Laboratory, Hangzhou, China}
\affiliation{I. Physikalisches Institut, Universit{\"a}t zu K{\"o}ln, Z{\"u}lpicher Stra{\ss}e 77, 50937 K{\"o}ln, Germany}

\author[0000-0002-9277-8025]{Thomas Möller}
\affiliation{I. Physikalisches Institut, Universit{\"a}t zu K{\"o}ln, Z{\"u}lpicher Stra{\ss}e 77, 50937 K{\"o}ln, Germany}

\author[0000-0002-8899-4673]{Xue-Jian Jiang}
\affiliation{Research Center for Astronomical computing, Zhejiang Laboratory, Hangzhou, China}

\author[0009-0008-0069-3501]{Zihao Song}
\affiliation{Research Center for Astronomical computing, Zhejiang Laboratory, Hangzhou, China}

\author[0009-0000-6108-2730]{Huaxi Chen}
\affiliation{Research Center for Astronomical computing, Zhejiang Laboratory, Hangzhou, China}

\author{Donghui Quan}
\affiliation{Research Center for Astronomical computing, Zhejiang Laboratory, Hangzhou, China}
\email{donghui.quan@zhejianglab.org}



\begin{abstract}
Interstellar molecules, which play an important role in astrochemistry, are identified using observed spectral lines. Despite the advent of spectral analysis tools in the past decade, the identification of spectral lines remains a tedious task that requires extensive manual intervention, preventing us from fully exploiting the vast amounts of data generated by large facilities such as ALMA. This study aims to address the aforementioned issue by developing a framework of automated line identification. We introduce a robust spectral fitting technique applicable for spectral line identification with minimal human supervision. Our method is assessed using published data from five line surveys of hot cores, including W51, Orion-KL, Sgr B2(M), and Sgr B2(N). By comparing the identified lines, our algorithm achieves an overall recall of $\sim 74\% - 93\%$, and an average precision of $\sim 78\% - 92\%$. Our code, named \textsc{spectuner}, is publicly available on GitHub.
\end{abstract}



\section{Introduction} \label{sec:intro}
Interstellar molecules, crucial for probing chemical and physical processes in the interstellar medium, are identified through their spectral lines. Current and future facilities, e.g. IRAM, ALMA, NOEMA, ng-VLA, and SKA, generate a vast amount of spectral line data annually. Analyzing the data presents a significant challenge.
\par
In the past decade, many tools have been developed to analyze and identify molecular spectral lines, including \textsc{weeds} \citep{2011A&A...526A..47M}, \textsc{cassis} \citep{2015sf2a.conf..313V}, \textsc{xclass} \citep{2017A&A...598A...7M}, \textsc{madcuba} \citep{2019A&A...631A.159M}, and \textsc{pyspeckit} \citep{2022AJ....163..291G}. Most of these tools treat spectral line identification as a curve fitting problem. In practice, spectral line identification involves excessive manual processes, including selecting molecules, making initial guesses for fitting, and checking the goodness of fit, which make spectral line identification a laborious task.
\par
The main goal of this study is to develop an automated line identification framework of interstellar species. Our algorithm, which builds on \textsc{xclass} \citep{2017A&A...598A...7M}, can automatically analyze an observed single-pointing spectrum and produce the best-fitting spectrum, associated parameters, and a table of identified lines with minor human intervention. Our algorithm is assessed using published line survey data of hot cores towards W51, Orion KL, Sgr B2(M) and Sgr B2(N).
\par
This paper is structured as follows. Section \ref{sec:method} describes our methodology. The line survey data used for evaluating our algorithm are introduced in Section \ref{sec:data}. Our results are presented in Section \ref{sec:app}. Finally, this work is summarized in Section \ref{sec:sum}.
\par
Our code, named \textsc{spectuner}, is publicly available on GitHub\footnote{\url{https://github.com/yqiuu/spectuner}} under a BSD 3-Clause License and version 1.0.0-beta is archived in Zenodo \citep{qiu_2024_14504040}.

\section{Methodology} \label{sec:method}

\begin{figure}
	\includegraphics[width=.4\columnwidth]{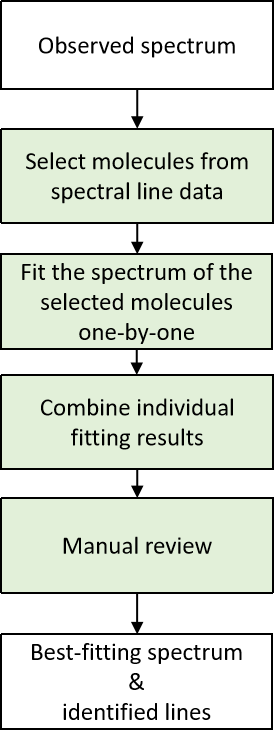}
	\centering
    \caption{Flowchart of the proposed automated line identification framework. An overview of the algorithm is given in Section \ref{sec:overview}.}
    \label{fig:flowchart}
\end{figure}

\subsection{Overview} \label{sec:overview}
Figure \ref{fig:flowchart} demonstrates the flowchart of our proposed framework for automated line identification. The input of the algorithm is an observed single-pointing spectrum, specified by its frequency and intensity. Our algorithm first queries a spectroscopic database in the observed frequency range and collects possible entries based on a user-defined list of molecules. Next, the algorithm sequentially performs fittings for all selected molecules, and then combines the fitting results using a greedy method. The combination process checks several criteria. Molecules that do not satisfy the criteria are put into a candidate list for manual review. Users can then decide whether to include a candidate. Ultimately, the algorithm outputs the best-fitting spectrum and parameters, along with a table listing the identified lines.
\par
We introduce the spectral line model and the spectroscopic database in Section \ref{sec:sl_model}. Our fitting techniques are detailed in Sections \ref{sec:opt} and \ref{sec:pm_loss}. The greedy method and the criteria for combining a fitting results are described in Section \ref{sec:combine}. A summary of the configuration parameters used in our algorithm is presented in Table \ref{tab:config}.

\subsection{The spectral line model} \label{sec:sl_model}
This study employs \textsc{xclass} \citep{2017A&A...598A...7M} to generate model spectra. The \textsc{xclass} code predicts model spectra by solving the one-dimensional radiative transfer equation for an isothermal object, assuming local thermodynamic equilibrium (LTE):
\begin{multline}
    T_\nu = I^\text{bg}_\nu - J^\text{CMB}_\nu \\
    + \sum_m \sum_c \eta(\theta^{m,c}) (J_\nu(T^{m,c}_\text{ex}) -  I^\text{bg}_\nu) (1 - e^{-\tau^{m,c}_\nu}), \label{eqn:inten}
\end{multline}
where the sums run over the indices $m$ for different species and their components $c$. Here, $I^\text{bg}_\nu$ is the background intensity, $J^\text{CMB}_\nu$ is the cosmic background intensity, $\eta(\theta^{m,c})$ is the beam filling factor, and $\theta^{m,c}$ is the source size, $T^{m,c}_\text{ex}$ is the excitation temperature. The source function $J_\nu(T)$ follows the black body radiation. Equation \ref{eqn:inten} assumes that the spectra of different species combine linearly.
\par
The optical depth $\tau^{m,c}_\nu$ is defined by
\begin{equation}
    \tau^{m,c}_\nu = \sum_t \tau^{m,c,t}_\nu,
\end{equation}
with
\begin{multline} 
    \tau^{m,c,t}_\nu = \frac{c^2}{8\pi\nu^2}  N^{m,c}_\text{tot} \phi(\nu;\nu^t, v^{m,c}_\text{LSR}, \Delta v^{m,c}) \\ \times \frac{g^t_\text{u} A^t_\text{ul}}{Q^m(T^{m,c}_\text{ex})} e^{-E^t_\text{l}/T^{m,c}_\text{ex}} \left(1 - e^{-h\nu^t/T^{m,c}_\text{ex}} \right), \label{eqn:tau}
\end{multline}
where $N_\text{tot}$ is the column density, $v_\text{LSR}$ is the velocity offset, $\Delta v$ is the velocity width, and $\phi(\nu)$ follows a Gaussian profile. The spectroscopic data used to compute the optical depth, i.e. $\nu^t$,$g^t_\text{u}$, $A^t_\text{ul}$, $E^t_\text{l}$, and $Q^m (T)$, are from the Virtual Atomic and Molecular Data Center (VAMDC) \citep{2016JMoSp.327...95E}, which contains entries from the Cologne Database for Molecular Spectroscopy (CDMS) \citep{2001A&A...370L..49M,2005JMoSt.742..215M} and the Jet Propulsion Laboratory (JPL) database \citep{1998JQSRT..60..883P}. Readers are referred to \citet{2017A&A...598A...7M} for a more detailed description for computing the model spectra.
\par
In this work, we fit the spectrum of a molecule, encompassing all possible states within the specified frequency range. Our code is capable of automatically selecting these possible states from the database. For a single component, the spectrum of one molecule of one state is characterized using five parameters: source size $\theta$, excitation temperature $T_\text{ex}$, column density $N_\text{tot}$, velocity width $\Delta v$, and velocity offset $v_\text{LSR}$. We assume that the source size, velocity width and velocity offset are shared among different states, while the other parameters are independent. This approach reduces the degrees of freedom, thereby simplifying the problem. 
\par
In addition, this study make the following assumptions:
\begin{enumerate}
    \item All molecules and isotopologues considered in this work are assumed to consist of only one component for simplicity.
    \item We do not account for any absorption features in the spectrum. Frequency channels below the background temperature are set to this temperature.
\end{enumerate}

\subsection{Optimization} \label{sec:opt}
This study utilizes particle swarm optimization (PSO) \citep[see e.g.][]{wang2018particle} for all fitting tasks. The algorithm, inspired by the social behavior of bird flocking, has a wide range of applications in both engineering and scientific fields. In the field of astronomy, PSO is applied to gravitational wave data analysis \citep{2014ApJ...795...96W}, calibration of semi-analytic galaxy formation models \citep{2015ApJ...801..139R}, and spectral energy distribution (SED) fitting \citep{2023MNRAS.519.2268Q}. Compared to the Levenberg–Marquardt algorithm, commonly used for spectral fitting, PSO offers several advantages. Firstly, PSO does not necessitate an initial guess from users. Instead, it selects initial points randomly within the specified parameter range. Secondly, the design of PSO balances exploration with exploitation, making it less prone to becoming trapped in local minima.
\par
In practice, we apply the PSO proposed by \cite{pso}. The initial positions are randomly drawn within the bounds specified in Table \ref{tab:bounds}. The optimization algorithm executes for a minimum of $N_{\rm min}$ iterations, with
\begin{equation}
    N_\text{min} = 100 + 5 \times (D - 5),
\end{equation}
where $D$ is the dimension of the problem. If the result fails to improve over 15 consecutive iterations, the optimization process will be halted. Moreover, a population size of 28 is adopted. This choice allows for running two tasks simultaneously on a single node of our high performance computing cluster. In general, a population size between 20 and 50 is preferred \citep{wang2018particle}.

\begin{table}
\centering
\caption{Summary of the bounds of the fitting parameters.}
\label{tab:bounds}
\begin{tabular}{cccc}
\hline
\hline
    Name & Unit & Scale & Bounds \\
\hline
    Source size $\theta$ & $''$ & log & 0.7 - 2.3 \\
    Excitation temperature $T_\text{ex}$ & K & linear & 10 - 400 \\
    Column density $N_\text{tot}$ & cm$^{-2}$ & log & 12 - 20 \\
    Velocity width $\Delta v$ & km/s & log & 0.0 - 1.5 \\
    Velocity offset $v_\text{LSR}$ & km/s & linear & -10 - 10 \\
\hline
\end{tabular}
\tablecomments{Initial positions for optimization are randomly sampled within these bounds, and the optimizer ensures that the parameters do not exceed the bounds.}
\end{table}

\begin{table*}
\centering
\caption{Summary of the configuration parameters of the proposed automated line identification algorithm.}
\label{tab:config}
\begin{tabular}{llll}
\hline
\hline
    Symbol & Value & Description & Section \\
\hline
\hline
    $N_\text{swarm}$ & 28$^a$ & Number of particles used in the particle swarm optimization. & Section \ref{sec:opt} \\
    $N_\text{trail}$ & 3 & Number of trails to run the optimization algorithm. & Section \ref{sec:opt} \\
\hline
\hline
    $R$ & 0.25 & Relative height to compute peak widths. & Section \ref{sec:pm_loss} \\
    $P$ & 4$\sigma$$^b$ & Required prominence of peaks. & Section \ref{sec:pm_loss} \\
\hline
\hline
    $S_X$ & $S_3 > 0.8$ & Threshold of the $X$-th largest score to combine a fitting result. & Section \ref{sec:combine} \\
    $S_\text{tot}$ & $S_\text{tot} > 2.7$  & Threshold of the total score to combine a fitting result. & Section \ref{sec:combine} \\
\hline
\end{tabular}
\tablecomments{$^a$Users may adjust this parameter from 20 to 50 based on the number of CPU cores to be used. $^b$This parameter should be provided by users. This work adopts four times the root mean square (RMS) noise. For practical applications, users should estimate the RMS noise by themselves.
}
\end{table*}

\subsection{The peak matching loss function} \label{sec:pm_loss}

\begin{figure}
	\includegraphics[width=\columnwidth]{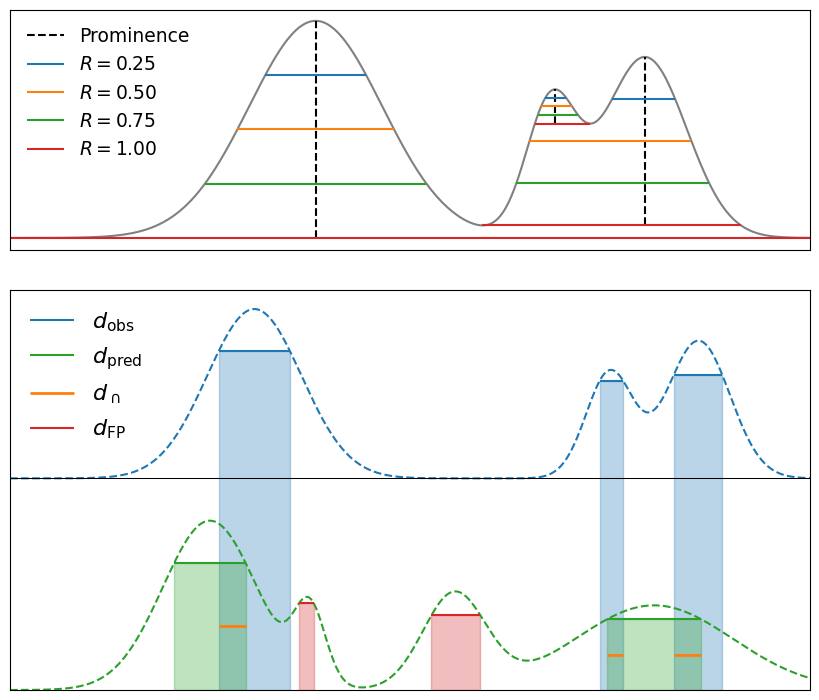}
	\centering
    \caption{Schematics of some definitions in our peak finding algorithm as introduced in Section \ref{sec:pm_loss}. In the upper panel, black dashed lines represent examples of prominence. Colored lines depict the peak widths estimated by the \texttt{peak\_widths} function in \textsc{scipy} using various input parameters $R$, defined in Equation \ref{eqn:h_width}. The lower panel illustrates the interval intersection algorithm assuming that the blue dashed line is an observed spectrum and the green dashed line is a model spectrum. The definitions of $d_\text{obs}$, $d_\text{pred}$, $d_\cap$, and $d_\text{FP}$ can also be found in the lower panel.}
    \label{fig:schematics}
\end{figure}

\begin{figure*}
	\includegraphics[width=\textwidth]{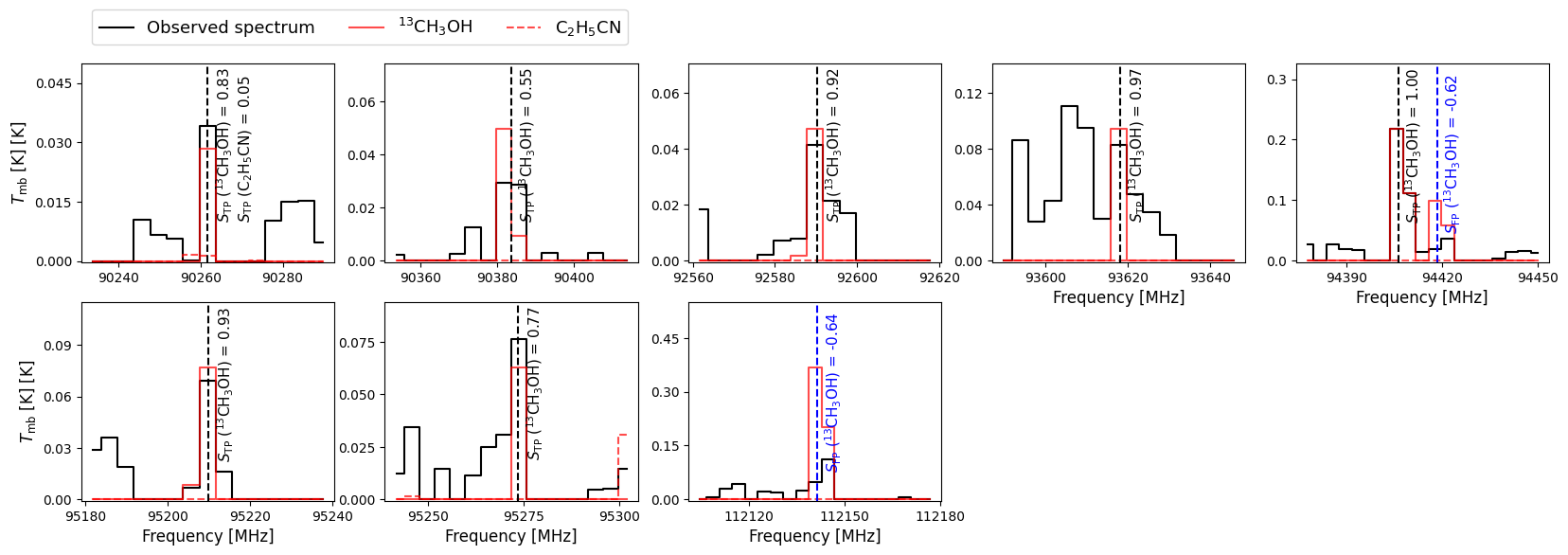}
	\centering
    \caption{Visualization of $S_\text{TP}^i$ and $S_\text{FP}^i$ defined in Equations \ref{eqn:s_start} to \ref{eqn:s_end}. The black solid lines illustrate the observed spectrum of \textit{W51-Mopra} (see Section \ref{sec:mopra}). The red lines show the best-fitting spectra of $^{13}$CH$_3$OH and C$_2$H$_5$CN obtained with the peak matching loss in the combined phase. The complete identification results for \textit{W51-Mopra} are presented in Section \ref{sec:app}. Each black (blue) vertical line represents a true (false) positive peak identified using our peak finding algorithm as described in Section \ref{sec:pm_loss}.}
    \label{fig:scores}
\end{figure*}

The loss function is a key component in optimization. Traditionally, the $\chi^2$ loss has been widely used for spectral fitting. However, we find that applying $\chi^2$ to perform fitting of a single molecule to observed spectra can be problematic and often leads to unsatisfactory results. The reason is that the $\chi^2$ loss is sensitive to outliers, i.e. features that cannot be described by the model. Since observed spectra, especially line-rich spectra, consist of spectral lines from a variety of molecules, they include many features that the model spectrum of a single molecule cannot describe.
\par
To address the above problem, this work proposes a peak matching loss function, which is able to robustly describe the distance between the observed and model spectra. Our idea is to only compare the spectra around the lines predicted by the spectral line model. The proposed loss function consists of three terms:
\begin{equation}
    \mathcal{L}_{\rm PM} = \mathcal{L}_{\rm MAE} + \sum_{\rm TPPs} \mathcal{L}_{\rm TP}  + \sum_{\rm FPPs} \mathcal{L}_{\rm FP}. \label{eqn:loss_tot}
\end{equation}
The first term is known as the mean absolute error (MAE),
which is defined as 
\begin{equation}
    \mathcal{L}_{\rm MAE} = \frac{1}{N} \sum_{i=0}^N \left| T^{\rm obs}_i - T^{\rm pred}_i \right|,
\end{equation}
where $N$ is the number of observed frequency channels. The MAE is commonly used in statistics and machine learning due to its robustness against outliers \citep{klebanov2009robust}. In addition, it can serve as a regularization term in regression models \citep{tibshirani1996regression}.

\par
To calculate the second and third terms, a peak finding algorithm must be applied to both observed and model spectra. This work adopts the \texttt{find\_peaks} and \texttt{peak\_widths} functions implemented in \textsc{scipy} \citep{2020SciPy-NMeth} to identify peaks. The algorithm includes a parameter, namely the required prominence, which is essential for our application. Prominence is a property of a peak, defined as the vertical distance from the peak height to the lowest contour line that does not enclose any higher peak. Examples of prominence are given in the upper panel of Figure \ref{fig:schematics}. The peak finder only identifies peaks with a prominence greater than the specified prominence threshold. In this study, the prominence threshold is set at four times the root mean square (RMS) noise. Additionally, our peak finder is designed to ignore peaks based on a predefined list of central frequencies. This function is useful for ignoring known radio recombination lines.
\par
After calling \texttt{find\_peaks}, we use \texttt{peak\_widths} to estimate the width of each peak. The widths are evaluated at $h_\text{width}$ defined as
\begin{equation} \label{eqn:h_width}
    h_\text{width} = h_\text{peak} - P \times R,
\end{equation}
where $h_\text{peak}$ is the peak height, $P$ is the prominence of the peak, and $R$ is a user-defined parameter. The choice of $R$ is subtle. As illustrated in the lower panel in Figure \ref{fig:schematics}, too large values of $R$ could result in the blending of several peaks, while too small values of $R$ might cause instability in calculating our peak matching loss function. Based on our experience, setting $R = 0.25$ is a practical choice.
\par
The objective of the aforementioned step is to characterize the peaks within a spectrum through an array of intervals defined by their central frequency and width. Once we obtain the intervals of both observed and model spectra, we apply an interval intersection algorithm to classify the peaks of the model spectrum as true and false positive peak. If an interval from the model spectrum intersects with an interval of the observed spectrum, the associated peak is labeled as a true positive peak (TPP), while if the condition is not satisfied, the corresponding peak is classified as a false positive peak (FPP). Illustrations of true and false positive peaks are presented in the lower panel of Figure \ref{fig:schematics}.
\par
The second term in Equation \ref{eqn:loss_tot} rewards true positive peaks (TPPs), which is defined as
\begin{equation}
    \mathcal{L}_\text{TP}=
        \begin{cases}
            D_\text{TP} - f_\text{dice} I^\text{obs}, & \, I^\text{pred} \geq I^\text{obs}, \\
            \min(D_\text{TP} - f_\text{dice} I^\text{obs}, 0), & \, I^\text{pred} < I^\text{obs}, \\
        \end{cases}
\end{equation}
with
\begin{align}
    f_\text{dice} &= \frac{2d_\cap}{d_\text{obs} + d_\text{pred}}, \\
    D_\text{TP} &= \frac{1}{d_\cap} \int_{d_\cap} \left|T^\text{obs}_\nu - T^\text{pred}_\nu \right| d\nu, \label{eqn:D_tp} \\
    I_\text{obs} &= \frac{1}{d_\text{obs}} \int_{d_\text{obs}} \left|T^\text{obs}_\nu - T^\text{back} \right| d\nu, \label{eqn:I_obs} \\
    I_\text{pred} &= \frac{1}{d_\text{pred}} \int_{d_\text{pred}} \left|T^\text{pred}_\nu - T^\text{back} \right| d\nu, \label{eqn:I_pred}
\end{align}
where $d_\cap$, $d_\text{obs}$, and $d_\text{pred}$ are shown in the lower panel of Figure \ref{fig:schematics}. The use of $f_\text{dice}$ has two effects. First, it smooths the transition between the cases with and without intersection. Secondly, it could measure the similarity in shape between two peaks. When computing the integrals in Equations \ref{eqn:D_tp}, \ref{eqn:I_obs}, and \ref{eqn:I_pred}, the trapezoidal rule is adopted. Within the interval, we only use channels given by the observed spectrum, without including any additional points, while linear interpolation is applied at the boundaries. Unlike a typical loss function such as the MAE, $\mathcal{L}_\text{TP}$ can produce negative values when the peaks in the observed and model spectra match well.
\par
The third term in Equation \ref{eqn:loss_tot} penalizes false positive peaks (FPPs). Its definition is given by
\begin{equation}
    \mathcal{L}_\text{FP} = D_\text{FP} W(\nu_\text{FP}),
\end{equation}
with
\begin{align}
    D_\text{FP} &= \min(D_1, D_2) \label{eqn:D_fp} \\
    D_1 &= \frac{1}{d_\text{FP}} \int_\text{FP} \max( T^\text{pred}_\nu -  T^\text{obs}_\nu, 0) d\nu \label{eqn:D_1} \\
    D_2 &= \frac{1}{d_\text{FP}} \int_\text{FP} \max( T^\text{pred}_\nu -  T^\text{back} - P, 0) d\nu \label{eqn:D_2} \\
    W(\nu) &= \begin{cases}
        \frac{\nu - \nu_\text{left}}{\nu_\text{mid} - \nu_\text{left}}, & \, \nu < \nu_\text{mid}, \\
        \frac{\nu_\text{right} - \nu}{\nu_\text{right} - \nu_\text{mid}}, & \,  \nu \geq \nu_\text{mid}, \\
    \end{cases} \\
    \nu_\text{mid} &= \frac{1}{2} (\nu_\text{left} + \nu_\text{right}),
\end{align}
where $P$ is the value of the prominence and $\nu_\text{FP}$ is the central frequency of the false positive peak. $\nu_\text{left}$ and $\nu_\text{right}$ are the central frequencies of the nearest observed peaks on the left and right sides respectively. The trapezoidal rule is adopted to compute Equations \ref{eqn:D_1} and \ref{eqn:D_2}. If a false positive peak is hidden in a broad observed peak, e.g. the left false positive peak in the lower panel of Figure \ref{fig:schematics}, there is no need to penalize it, and therefore we introduce a maximum operation in Equations \ref{eqn:D_1} and \ref{eqn:D_2}. The use of $D_2$ is to smooth the transition between the cases if a local maximum is identified and not identified as a peak. $W(\nu)$ allows to reduce $\mathcal{L}_\text{FP}$ by shifting the central frequency in addition to decreasing $T^\text{pred}_\nu$.

\subsection{Combining individual fitting results} \label{sec:combine}
This work proposes a greedy algorithm to combine individual fitting results (see Algorithm \ref{alg:combine}). In the algorithm, a "pack" consists of a molecule list, model parameters and a model spectrum, which can initiate a new fitting. The algorithm input is a list of packs, each constructed from individual fitting  results. The basic idea of the algorithm is that fitting results with lower values can better match more features in the observed spectrum. The algorithm initially sorts the individual fitting results by their loss values, then combines them sequentially, beginning with the one having the lowest loss value. A new optimization will be performed if there is line blending. The algorithm outputs a combined result and fitting results of some candidate molecules which do not satisfy the criteria described in the following paragraph. 
\par
The proposed algorithm adopts two scores to decide whether an individual fitting result can be incorporated. First, the total score is defined as
\begin{equation}
    S_\text{tot}^i = \sum_\text{TPPs} S_\text{TP}^i + \sum_\text{FPPs} S_\text{FP}^i,
\end{equation}
with
\begin{align}
    S_\text{tot}^i &= \sum_\text{TPPs} S_\text{TP}^i + \sum_\text{FPPs} S_\text{FP}^i, \\
    S_\text{TP}^i &= F_\text{TP}^i \max \left( 1 - \frac{D_\text{TP}^i}{I_\cap^i}, 0  \right), \label{eqn:s_start} \\
    F_\text{TP}^i &= \frac{I_\cap^i}{\sum_{i}^{'} I_\cap^i}, \\
    I_\cap^i &= \frac{1}{d_\cap} \int_{d_\cap} \left|T^{\text{pred}, i}_\nu - T^\text{back} \right| d\nu, \\
    S_\text{FP}^i &= -F_\text{FP}^i \frac{D_\text{FP}^i}{I_\text{pred}^i}, \\
    F_\text{FP}^i &= \frac{I_\text{pred}^i}{\sum_{i}^{'} I_\text{pred}^i}, \label{eqn:s_end} 
\end{align}
where $I_\text{pred}^i$, $D_\text{TP}^i$, and $D_\text{FP}^i$ are defined in Equations \ref{eqn:I_pred}, \ref{eqn:D_tp}, and \ref{eqn:D_fp}. These quantities here are computed for each molecule. We note that a true (or false) positive peak can be contributed by multiple molecules, and we use $F_\text{TP}^i$ and $F_\text{FP}^i$ to take into account the different contributions. In the calculation of $F_\text{TP}^i$ and $F_\text{FP}^i$, contributions less than $5\%$ are excluded beforehand, resulting in renormalized fractions. Figure \ref{fig:scores} provides a visualization of $S_\text{TP}^i$ and $S_\text{FP}^i$. 
\par
We further define $S_X^i$ as the $X$-th largest value among $S^i_\text{TP}$ of all true positive peaks. If the number of true positive peaks is smaller than $X$, $S_X^i$ is set to zero.
\par
Our criteria to combine a fitting result are 
\begin{align}
    S_\text{3}^i > 0.8 \; \text{and} \;  S_\text{tot}^i > 2.7.
\end{align}
Satisfying these criteria implies that at least three lines are matched. Specifically, the lower limit of $S_\text{3}^i$ ensures that the model spectrum has at least three matched lines with a score exceeding 0.8, while $S_\text{tot}^i$ imposes additional constraints on the average score and accounts for false positive peaks. We note that these values are purely empirical but lead to reasonable results. Users have the flexibility to adjust the criteria according to their specific applications.
\par
We point out that a more straightforward method for combining all individual fitting results could involve simultaneously fitting the spectrum of all molecules. However, this method is not adopted in this study. The large number of molecules involved in the fitting task results in an excessive number of free parameters ($\sim$ 100 to 500), posing a significant challenge. \cite{2024ApJ...965...14E} conducted a spectral line analysis for NGC 6334I, who simultaneously fit the spectrum of 21 molecules. They fixed the excitation temperature and line width of each molecule to reduce the degrees of freedom. Furthermore, due to the intricacy of the spectral lines, there are various choices for combining the individual fitting results, and our algorithm offers an interpretable solution.

\SetKwComment{Comment}{//}{}
\RestyleAlgo{ruled}
\begin{algorithm*}
\caption{Combining individual fitting results} \label{alg:combine}

%\BlankLine
\# The algorithm is introduced in Section \ref{sec:combine}, including the criteria to accept a result.  \\
\KwIn{$pack\_list$}
\KwOut{$pack\_combined$, $cand\_list$}
Sort $pack\_list$ by loss in ascending order\;
Set $cand\_list$ as an empty list\;
$pack\_combined \gets None$\;
\# Determine the first pack that can be merged. \\
$i\_pack \gets 0 $\;
\For{$i = 0, ..., \text{len}(pack\_list)$} {
    $pack \gets pack\_list[i]$\;
    \rm Estimate the $scores$ for $pack$\;
    \eIf{\rm the $scores$ satisfy the criteria}{
        $pack\_combined \gets pack$\;
        break\;
    }{\rm Put $pack$ into $cand\_list$\;}
    $i\_pack \gets i\_pack + 1$\;
}
\BlankLine
\# Merge the remaining packs. \\
\For{$i = i\_pack + 1, ..., \text{len}(pack\_list)$} {
    $pack \gets pack\_list[i]$\;
    \If{$pack$ \rm has overlapped lines with $pack\_combined$}{
        Use $pack$ to start a new fitting, including the best-fitting parameters of $pack$ in the initialization of the optimizer and setting the spectrum of $pack\_combined$ as background\;
        Overwrite $pack$ using the new fitting result\;
    }
    Combine $pack\_combined$ with $pack$ to estimate the $scores$ for $pack$\;
    \eIf{\rm the $scores$ satisfy the criteria}{
        Merge $pack$ into $pack\_combined$\;
    }{\rm Put $pack$ into $cand\_list$\;}
}
\BlankLine
\# Perform a new fitting for all candidates. \\
\For{$i = 0, ..., \text{len}(cand\_list)$} {
    $pack \gets cand\_list[i]$\;
    \If{$pack$ \rm has overlapped lines with $pack\_combined$}{
        Use $pack$ to start a new fitting and set the spectrum of $pack\_combined$ as background\;
        Overwrite $pack$ using the new fitting result\;
        $cand\_list[i] \gets pack$\;
    }
}
\end{algorithm*}

\section{Data} \label{sec:data}
In this work, we assess our algorithm for automated line identification using data from five published line surveys.

\subsection{W51-Mopra} \label{sec:mopra}
We utilize the published data from \cite{2017ApJ...845..116W}, who conducted a line survey towards the W51 region using the Mopra 22 m telescope. The survey covers the frequency ranges from 85.2 - 101.1 GHz and 107.0 - 114.9 GHz. The spectral data originate from the hot cores e1/e2, located at galactic coordinate ($l$, $b$) = (49$^{\circ}$.4898, -0$^{\circ}$.3874). \cite{2017ApJ...845..116W} identified 234 emission lines. By matching the spectral line databases, namely CDMS and JPL, they identified 31 molecules and 18 isotopologues. No parameter estimation is provided for the spectrum of the hot cores e1/e2.
\par
We adopt the RMS noise values of 20 mK and 70 mK for 85.2 - 101.1 GHz and 107.0 - 114.9 GHz respectively, as indicated in Figure 2 of \cite{2017ApJ...845..116W}. These values serve as inputs for our peak finding algorithm. In addition, their table of identified lines includes six radio recombination lines, which are excluded by our peak finder.

\subsection{OrionKL-GBT} \label{sec:bgt}
This study relies on the data published by \citet{2015AJ....149..162F}, who conducted a line survey towards Orion KL. The observations were carried out using the Green Bank Telescope (GBT) and were centered on the celestial coordinates R.A. (J2000) 05:35:14, decl. (J2000) 05:22:27.5. The frequency range covered by the survey spans from 67.0 GHz to 93.6 GHz. \citet{2015AJ....149..162F} only identified lines with intensities brighter than 1 K. By matching the Splatalogue database, a total of 140 emission lines were identified, with no parameter estimation provided.
\par
As input for our peak finder, we adopt three RMS noise levels: 66.5 mK for 67.0 GHz to 70.0 GHz, 31.6 mK for 70.0 GHz to 85 GHz, and 41.2 mK for 85.0 GHz to 93.6 GHz. These values are estimated from their Table 1. \citet{2015AJ....149..162F} identified three lines from atmospheric O$_2$, which are ignored by our peak finder.

\subsection{OrionKL-Tianma} \label{sec:tianma}
We incorporate the data from \citet{2022ApJS..263...13L}, who carried out a line survey targeting the Orion KL region using the Tianma 65 m radio telescope. The spectral data are centered on the celestial coordinates R.A. (J2000) 05:35:14.55, decl. (J2000) -05:22:31.0. The frequency coverage spans from 34.5 GHz to 50.0 GHz. A total of 597 emission lines were identified, including 177 classified as radio recombination lines. \citet{2022ApJS..263...13L} carried out a spectral line fitting analysis for the molecular lines using a one-dimensional LTE model. Their spectral line model employs equations similar to Equations \ref{eqn:inten} to \ref{eqn:tau}. In their analysis, they adopted fixed values for the source size and excitation temperature, while fitting the column density, velocity width, and velocity offset.
\par
For the peak finder, we adopt two RMS noise values, at 4 mK and 7 mK for 34.5 GHz - 48.7 GHz and 48.7 GHz - 50.0 GHz respectively. We estimate these values from Figure 3 in \citet{2022ApJS..263...13L}. All radio recombination lines identified by \citet{2022ApJS..263...13L} are ignored by our peak finder.

\subsection{SgrB2M-IRAM and SgrB2N-IRAM} \label{sec:b2}
The most complex test data utilized in this study originate from \citet{2013A&A...559A..47B}, who presented a line identification analysis for both Sgr B2(M) and B2(N) within the frequency range of 80.0 GHz to 116.0 GHz. The observations were conducted using the IRAM 30 m telescope, targeting the positions at R.A. (J2000) 17:47:20s.4, decl. (J2000) -28:23:07.0 for Sgr B2(M), and R.A. (J2000) 17:47:20.0 and decl. (J2000) -28:22:19.0 for and B2(N), respectively. These sources are line rich. \citet{2013A&A...559A..47B} identified 945 and 3675 emission lines in Sgr B2(M) and B2(N) respectively. They fit the observed spectra using the \textsc{xclass} spectral model to estimate the source size, column density, excitation temperature, velocity width and velocity offset. The parameters were adjusted through visual inspection.
\par
In terms of the RMS noise required by our peak finder, we use four values as listed in Table 1 in \citet{2013A&A...559A..47B}.

\begin{table*}
\centering
\caption{Summary of the line survey data used in this study and the recalls attained using our automated line identification algorithm.}
\label{tab:result}
\begin{tabular}{cccccccc}
\hline
\hline
Name & Band & Line density$^a$ & Recall$^b$ ($\chi^2$) & Recall$^b$ (PM) & Prec.$^c$ ($\chi^2$) & Prec.$^c$ (PM) \\
 & (GHz) & (GHz$^{-1}$) &  &  \\
\hline
\textit{W51-Mopra} & 85 - 101, 107 - 115 & 13.8 & 61\% (93/152) & 84\% (127/152) & 73\% & 90\% \\
\textit{OrionKL-GBT} & 67 - 94 & 20.3 & 82\% (86/105) & 93\% (98/105) & 86\% & 89\% \\
\textit{OrionKL-Tianma} & 35 - 50 & 30.5 & 57\% (136/237) & 83\% (196/237) & 62\% & 89\% \\
\textit{SgrB2M-IRAM} & 80 - 116 & 38.0 & 39\% (181/463) & 78\% (363/463) & 56\% & 92\% \\
\textit{SgrB2N-IRAM} & 80 - 116 & 59.2 & 31\% (329/1055) & 74\% (784/1055) & 51\% & 78\% \\
\hline
\end{tabular}
\tablecomments{$^a$The total number of lines for estimating line density is obtained using our peak finder described in Section \ref{sec:pm_loss}. $^b$The overall recall is defined in Section \ref{sec:metrics}. $^c$ The average precision is defined in Section \ref{sec:metrics}.}
\end{table*}

\begin{figure*}
	\includegraphics[width=\textwidth]{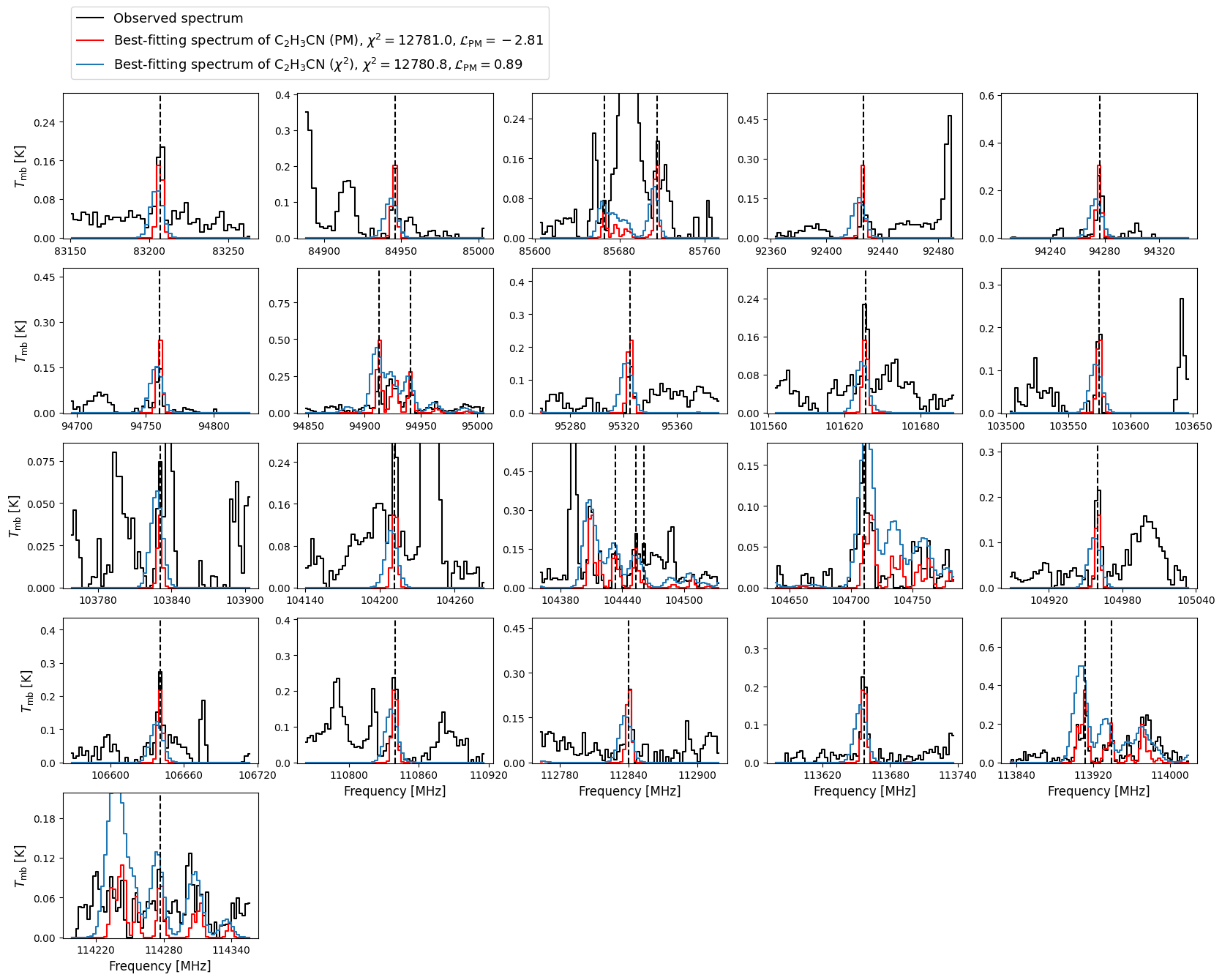}
	\centering
    \caption{Comparison of the best-fitting spectra using different loss functions. The black solid line illustrates the observed spectrum of \textit{SgrB2M-IRAM} (see Section \ref{sec:b2}), and blue dashed lines represent the spectral lines of C$_2$H$_3$CN identified by \citet{2013A&A...559A..47B}. The blue and red lines demonstrate the best-fitting results achieved with the $\chi^2$ and peak matching loss functions respectively. The peak matching loss function is defined in Section \ref{sec:pm_loss}. We compute and display both the $\chi^2$ and peak matching loss values corresponding to each best-fitting spectrum. These values imply that using the $\chi^2$ loss is inappropriate for spectral fitting of a single molecule.}
    \label{fig:chi2_pm}
\end{figure*}

\section{Applying to line surveys} \label{sec:app}
In this study, for speed reason, we downsample all observed spectra by averaging every two adjacent channels. The downsampling is repeated three times. Accordingly, the number of channels in the observed spectra is reduced by a factor of eight, and the RMS noise values are divided by $\sqrt{8}$ before being input into the peak finding algorithm. Secondly, for simplicity, this work neglects any absorption in the observed spectra. Given that the baselines of all the observed spectra used in this work have been calibrated to zero by the original studies, our correction is straightforward. Specifically, any channels displaying a negative value are reset to zero. In addition, the continuum spectra are not taken into account in the fitting processes. This approach may have negligible effects if the excitation temperature of the molecule is significantly higher than that of the continuum. Conversely, for molecules whose excitation temperature is comparable to the background temperature, neglecting the continuum spectra could affect the estimated excitation temperature and column density.
\par
Furthermore, our algorithm requires a molecule list as input. Since this work focuses on whether our method can reproduce the identification results of the previous studies rather than identifying new molecules, we use the union of the molecules and isotopologues identified in all the five line surveys as the input list, which contains 105 species. For practical use, our code provides a list of commonly observed molecules, e.g. methanol (CH$_3$OH), methyl formate (CH$_3$OCHO) and cyanoacetylene (HCCCN). Users can customize the molecule list to suit their specific applications. Users may also refer to \cite{2022ApJS..259...30M} for molecules that have been detected in different sources. 

\subsection{Metrics} \label{sec:metrics}
In this work, we use recall and precision to compare our identification results with the published line survey data. Recall measures the ability of our algorithm to find all spectral lines identified by the previous studies, while precision measures the accuracy of our line assignments.
\par
To compute the metrics, for each line survey, we prepare a line table for comparison that contains only lines identified by our peak finder in both the final predicted spectrum and the downsampled observed spectrum. The downsampling introduced earlier may result in deviations of the peak frequencies. To match the lines between the predicted and observed spectra, we use the peak widths determined by the \textsc{scipy} function \texttt{peak\_widths} (see Section \ref{sec:pm_loss}) as the tolerance criteria. When constructing the line tables for comparison, the following points are taken into account:
\begin{enumerate}
    \item A peak in the observed spectrum may be labeled by multiple transitions from the same molecule. In such cases, these transitions are treated as a single line.
    \item If a line in the observed spectrum is labeled by multiple different species, that line is excluded. Since the original study does not provide the contributions for each species, comparing such lines is tricky. In contrast, if a line in the predicted spectrum is labeled by multiple different species, we label the line with the one that has the largest contribution.
    \item The line surveys may use different spectral line databases, and to be consistent, we only include lines that are present in the spectral line database used in this work.
    \item Radio recombination lines identified in the observed spectrum are excluded.
\end{enumerate}

\par
We define the recall and precision metrics as
\begin{align}
    \text{Recall} &= \frac{N_\text{match}}{N_\text{ref}}, \\
    \text{Precision} &= \frac{N_\text{match}}{N_\text{pred}}.
\end{align}
For a given molecule, $N_\text{ref}$ is the number of lines identified by the referred study, $N_\text{pred}$ is the number of lines labeled as that molecule by our algorithm, and $N_\text{match}$ is the number of lines that match between the referred study and our algorithm. For example, if the referred study identifies 10 methanol lines, and our algorithm labels 9 methanol lines, where 8 lines match those in the referred study, and 1 line is labeled as a different molecule, then the recall of methanol is $8/10 = 0.8$, and the precision of methanol is $8/9 \approx 0.89$. In addition, when counting $N_\text{ref}$, $N_\text{pred}$, and $N_\text{match}$, we only consider those in the line table for comparison as described in the previous paragraph.
\par
Furthermore, we compute the overall recall for each line survey by dividing the total number of lines identified by our algorithm by the number of lines in the table for comparison. We also calculate the average precision for each line survey, which is the average of the precisions for all molecules.

\subsection{Results}
Table \ref{tab:result} presents a comparison of the overall recall and average precision obtained using the $\chi^2$ and peak matching loss functions for each line survey. Across all the line survey data, the peak matching loss function consistently outperforms the $\chi^2$ loss function in terms of both the recall and precision values.
\par
We illustrate the issue with using the $\chi^2$ loss function through an example. In Figure \ref{fig:chi2_pm}, we present the fitting results of C$_2$H$_3$CN during the individual fitting phase for \textit{SrgB2N-IRAM}. The results based on the peak matching and $\chi^2$ loss functions are represented by red and blue lines respectively. It is evident that the red line gives a better result than the blue line. However, upon calculating the $\chi^2$ and peak matching loss values for both fits, we find that the $\chi^2$ value of the blue line ($\chi^2$ = 12780.8) is smaller than the red line ($\chi^2$ = 12781.0). In other words, if the $\chi^2$ loss is adopted, the algorithm will consider the blue line as a better result than the red line, irrespective of optimization methods. Such result is unexpected. In contrast, the proposed peak matching loss assigns a lower value to the red line ($\mathcal{L}_\text{PM} = -2.81$) than the blue line ($\mathcal{L}_\text{PM} = 0.89$), which appropriately reflects the variance between the observed and fitting spectra.
\par
The results discussed above align with the practice of using $\chi^2$, where careful settings of initial guess and search space are necessary. For example, the result based on the $\chi^2$ loss can be improved by imposing an upper limit of the line width. These findings imply that the $\chi^2$ loss function may not be ideal for an automated spectral line fitting framework, where we aim for minimal manual settings, whereas the proposed peak matching loss function appears to be a superior alternative.

\subsubsection{W51-Mopra} \label{sec:res_mopra}
\begin{figure*}
	\includegraphics[width=\textwidth]{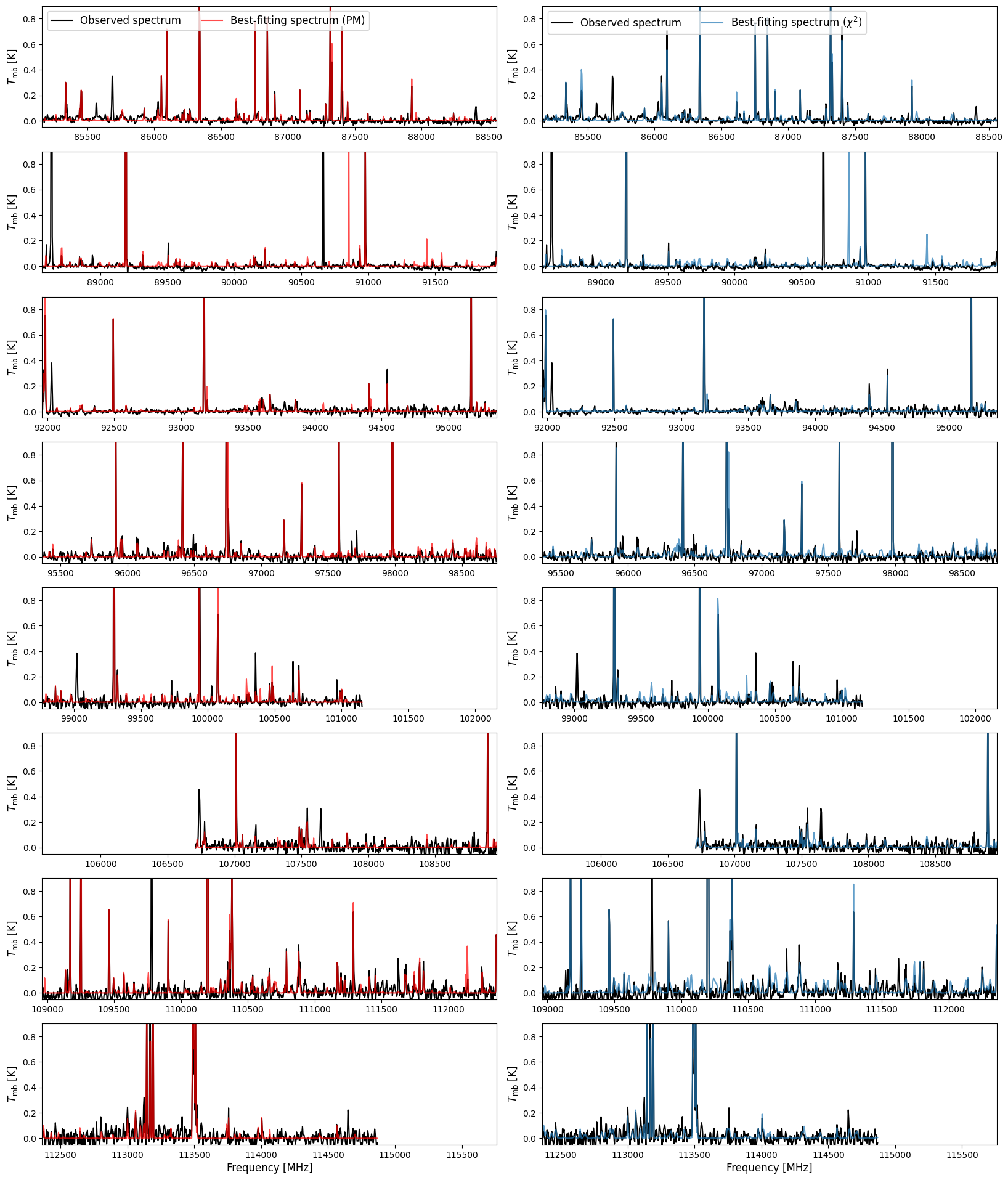}
	\centering
    \caption{Best fitting results for \textit{W51-Mopra}. Black lines represent the observed spectrum. Left and right panels show the results using the peak matching and $\chi^2$ loss functions respectively. The peak matching loss function is described in Section \ref{sec:pm_loss}.}
    \label{fig:spec_watanabe17}
\end{figure*}

\begin{figure*}
	\includegraphics[width=\textwidth]{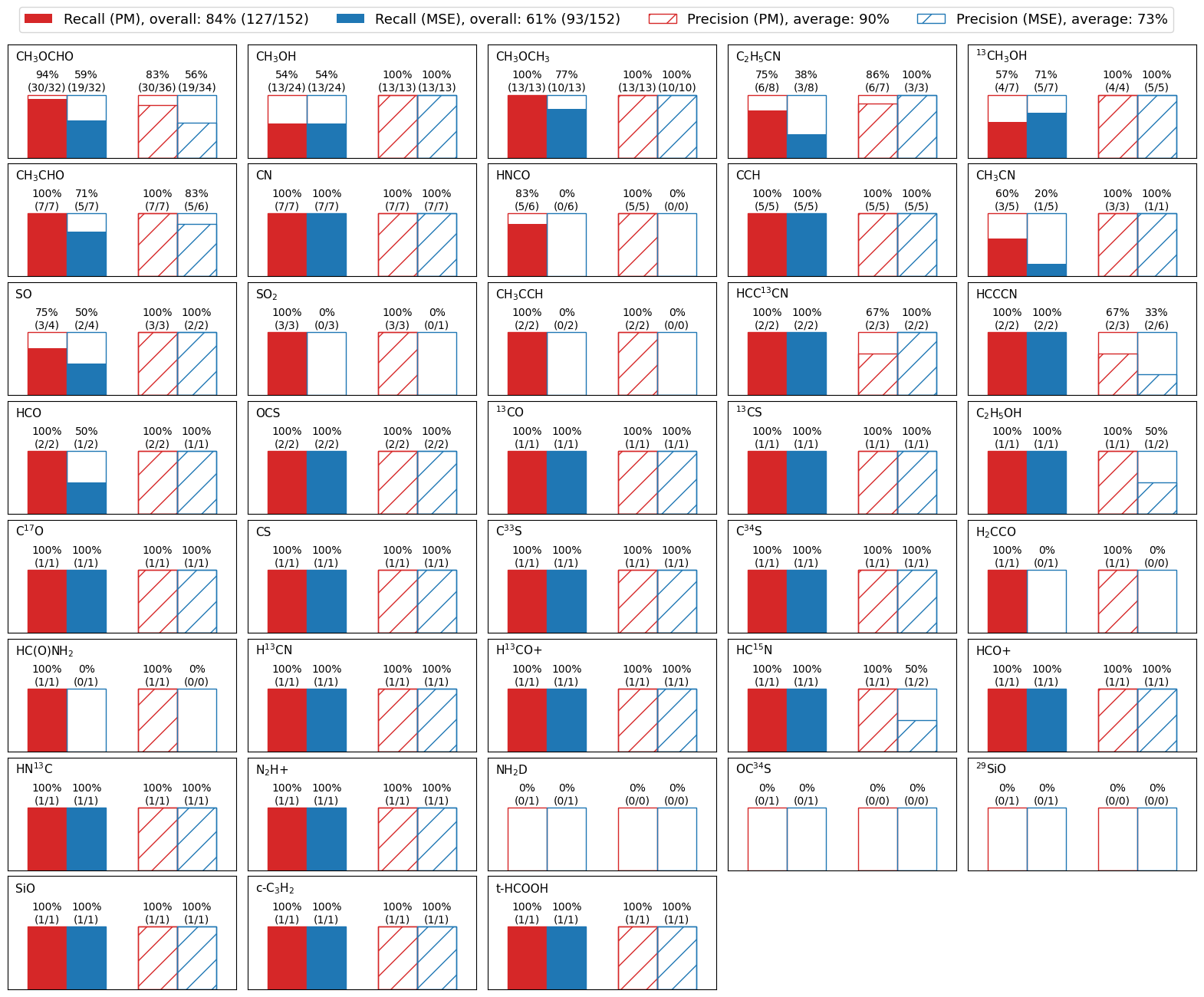}
	\centering
    \caption{Recall and precision of species identified in \textit{W51-Mopra}. The red and blue bars represent the results using the peak matching and $\chi^2$ loss functions respectively. Definitions of recall and precision are provided in Section \ref{sec:metrics}.}
    \label{fig:recall_watanabe17}
\end{figure*}

\begin{figure*}
	\includegraphics[width=\textwidth]{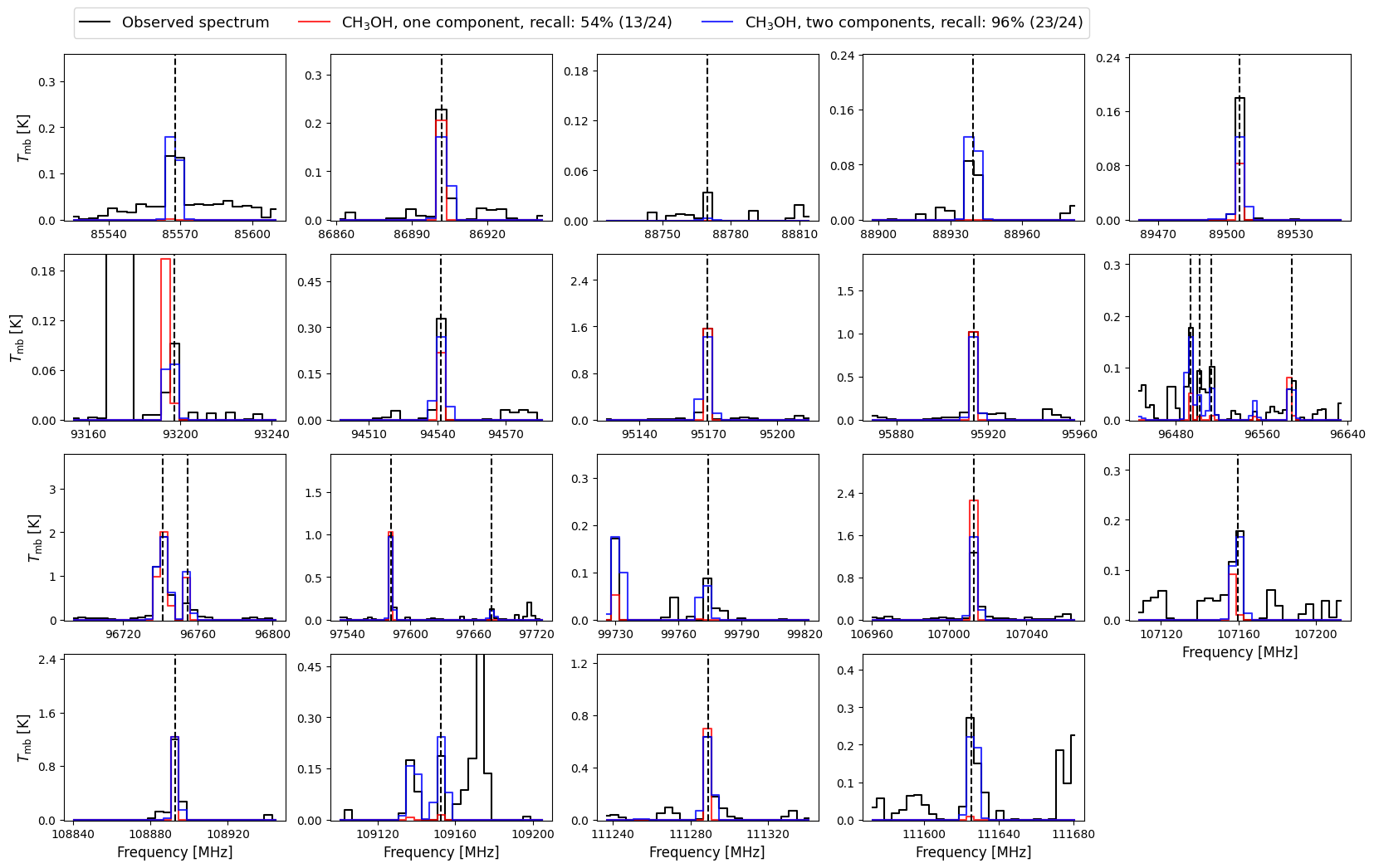}
	\centering
    \caption{Spectral lines of methanol in \textit{W51-Mopra}. The black solid lines show the observed spectrum from \citet{2017ApJ...845..116W}, and the black vertical lines are spectral lines of methanol identified by \citet{2017ApJ...845..116W}. The red and blue lines show the fitting results using one and two physical components respectively. The recall of each fitting result is indicated in the legend. These plots imply that the spectrum of methanol can be better modeled using two physical components.}
    \label{fig:watanabe17_CH3OH}
\end{figure*}

We illustrate the best-fitting spectrum for \textit{W51-Mopra} in Figure \ref{fig:spec_watanabe17}, and show the recalls and precisions in Figure \ref{fig:recall_watanabe17}. The overall recall improves from 61\% using the $\chi^2$ loss to 84\% using the peak matching loss. The average precision also increases from 82\% to 93\%. The recall of methanol stands relatively low, at 54\%. Our analysis suggests that the spectrum of methanol can be better described by more than one component. As demonstrated in Figure \ref{fig:watanabe17_CH3OH}, by adding an additional component, the best-fitting spectrum of methanol is improved, and the recall increases to 96\%.

\subsubsection{OrionKL-GBT}
\begin{figure*}
	\includegraphics[width=\textwidth]{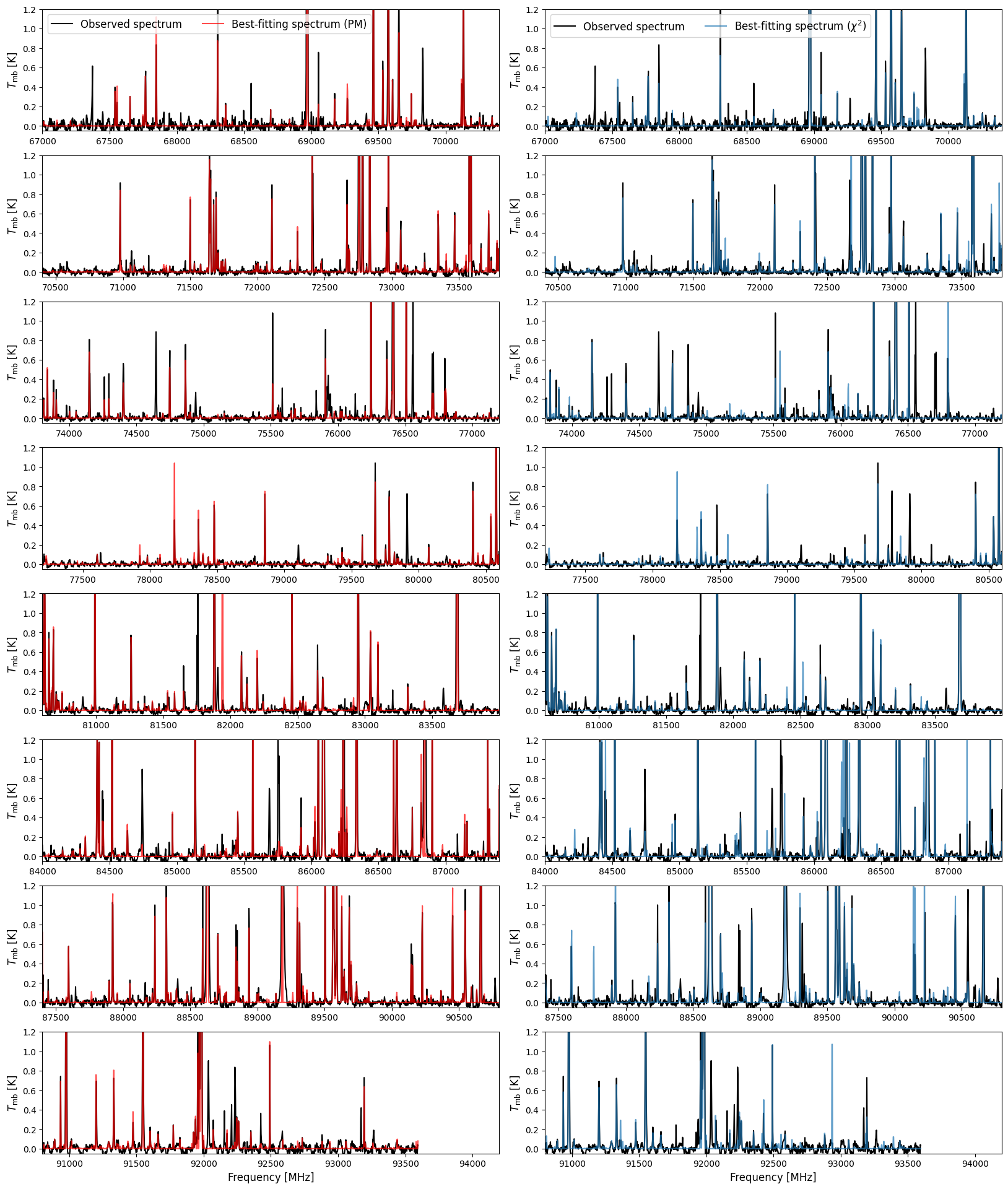}
	\centering
    \caption{Best fitting results for \textit{OrionKL-GBT}. Black lines represent the observed spectrum. Left and right panels show the results using the peak matching and $\chi^2$ loss functions respectively. The peak matching loss function is described in Section \ref{sec:pm_loss}.}
    \label{fig:spec_frayer15}
\end{figure*}

\begin{figure*}
	\includegraphics[width=\textwidth]{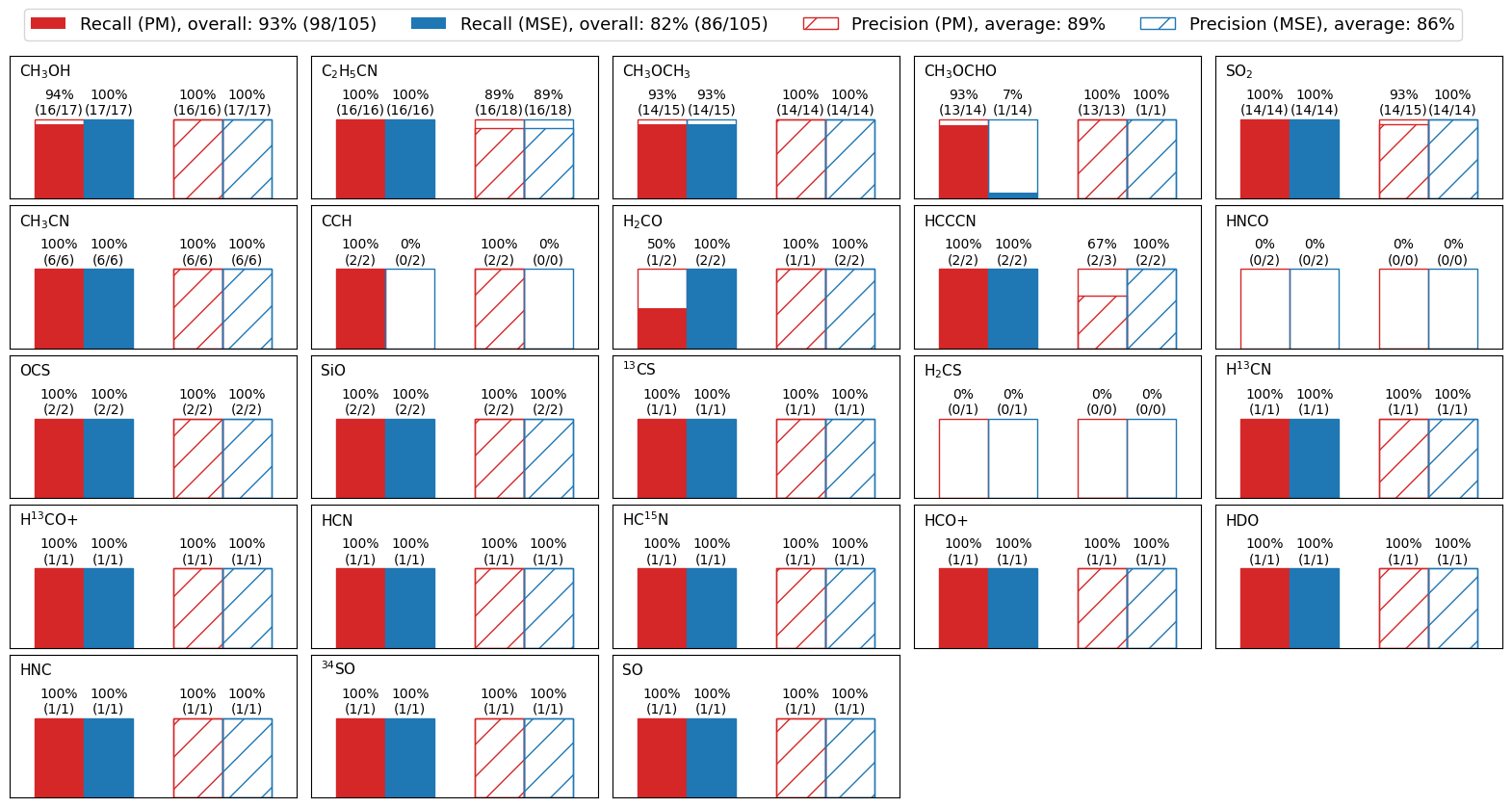}
	\centering
    \caption{Recall and precision of species identified in \textit{OrionKL-GBT}. The red and blue bars represent the results using the peak matching and $\chi^2$ loss functions respectively. Definitions of recall and precision are provided in Section \ref{sec:metrics}.}
    \label{fig:recall_frayer}
\end{figure*}

\begin{figure*}
	\includegraphics[width=0.65\textwidth]{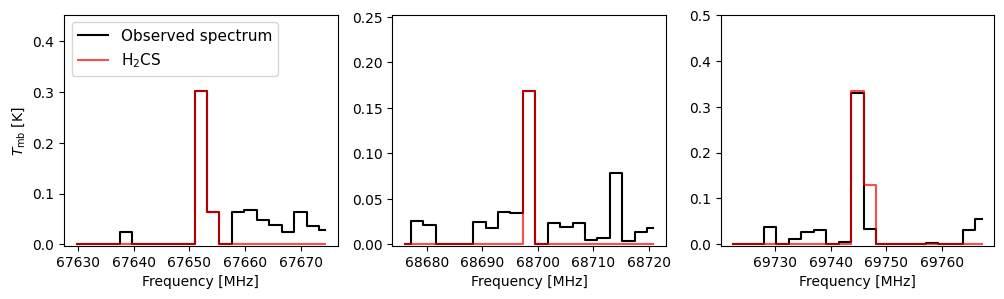}
	\centering
    \caption{Identified lines of H$_2$CS in \textit{OrionKL-GBT}. The red lines show our best-fitting spectrum of H$_2$CS using the peak matching loss function. The result implies a different velocity component of H$_2$CS in addition to the one identified by \citet{2015AJ....149..162F}.}
    \label{fig:frayer15_H2CS}
\end{figure*}

\begin{figure*}
	\includegraphics[width=0.43\textwidth]{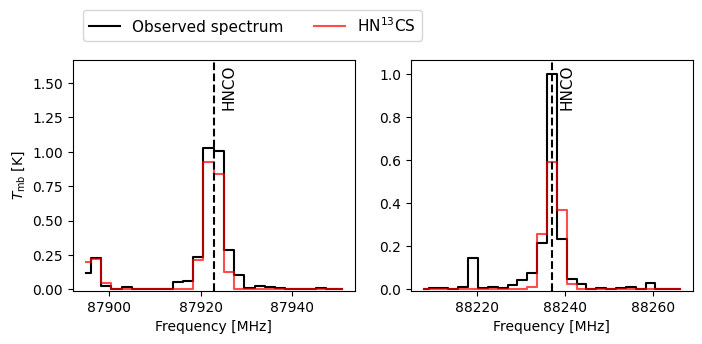}
	\centering
    \caption{Incorrect line assignment of HNCO in \textit{OrionKL-GBT} by our algorithm. The black vertical dashed lines indicate the two HNCO lines identified by \citet{2015AJ....149..162F}. Our algorithm incorrectly assign them to HN$^{13}$CO as indicated by the red lines, which show the best-fitting spectrum of HN$^{13}$CO using the peak matching loss function.}
    \label{fig:frayer15_HNCO}
\end{figure*}

For \textit{OrionKL-GBT}, we present the best-fitting spectra in Figure \ref{fig:spec_frayer15}. The recall and precision metrics of the molecules are demonstrated in Figure \ref{fig:recall_frayer}.\citet{2015AJ....149..162F} identified only those lines that are brighter than 1 K, which is significantly above the RMS noise. Therefore, we obtain the best results on \textit{OrionKL-GBT}, with a recall of 93 \% using the peak matching loss. Our algorithm fails to identify one line for thioformaldehyde (H$_2$CS). Our analysis suggests that the emission from thioformaldehyde involves two velocity components. Besides the one line identified by \citet{2015AJ....149..162F}, our method has detected a different velocity component comprising three lines, as illustrated in Figure \ref{fig:frayer15_H2CS}. Additionally, our algorithm misidentified isocyanic acid (HNCO), resulting from an incorrect assignment. As shown in Figure \ref{fig:frayer15_HNCO}, the two HNCO lines identified by \citet{2015AJ....149..162F} were incorrectly assigned to HN$^{13}$CO in our result.

\subsubsection{OrionKL-Tianma}
\begin{figure*}
	\includegraphics[width=\textwidth]{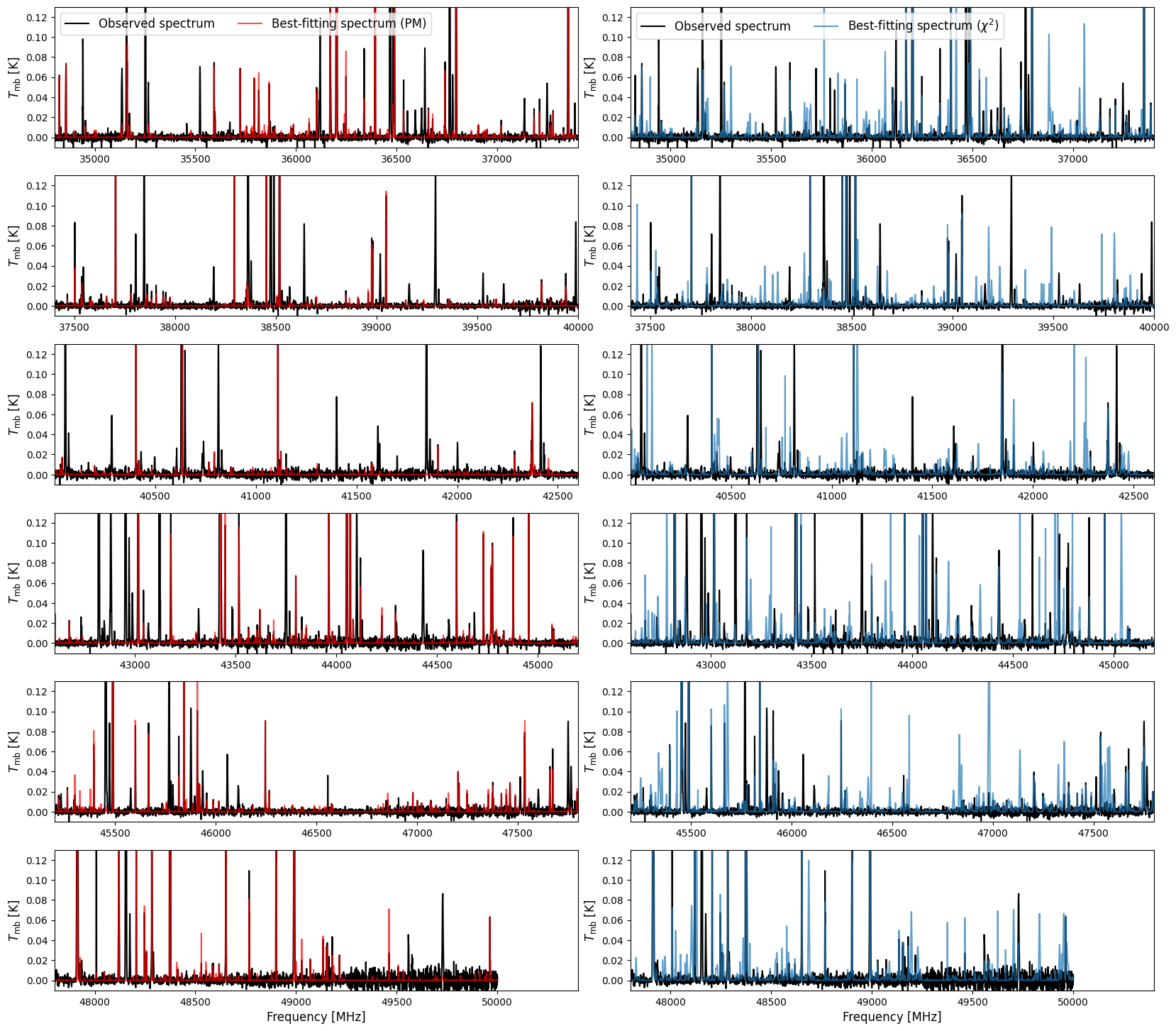}
	\centering
    \caption{Best fitting results for \textit{OrionKL-Tianma}. Black lines represent the observed spectrum. Left and right panels show the results using the peak matching and $\chi^2$ loss functions respectively. The line survey data are introduced in Section \ref{sec:tianma}. The peak matching loss function is described in Section \ref{sec:pm_loss}.}
    \label{fig:spec_tianma}
\end{figure*}

\begin{figure*}
	\includegraphics[width=\textwidth]{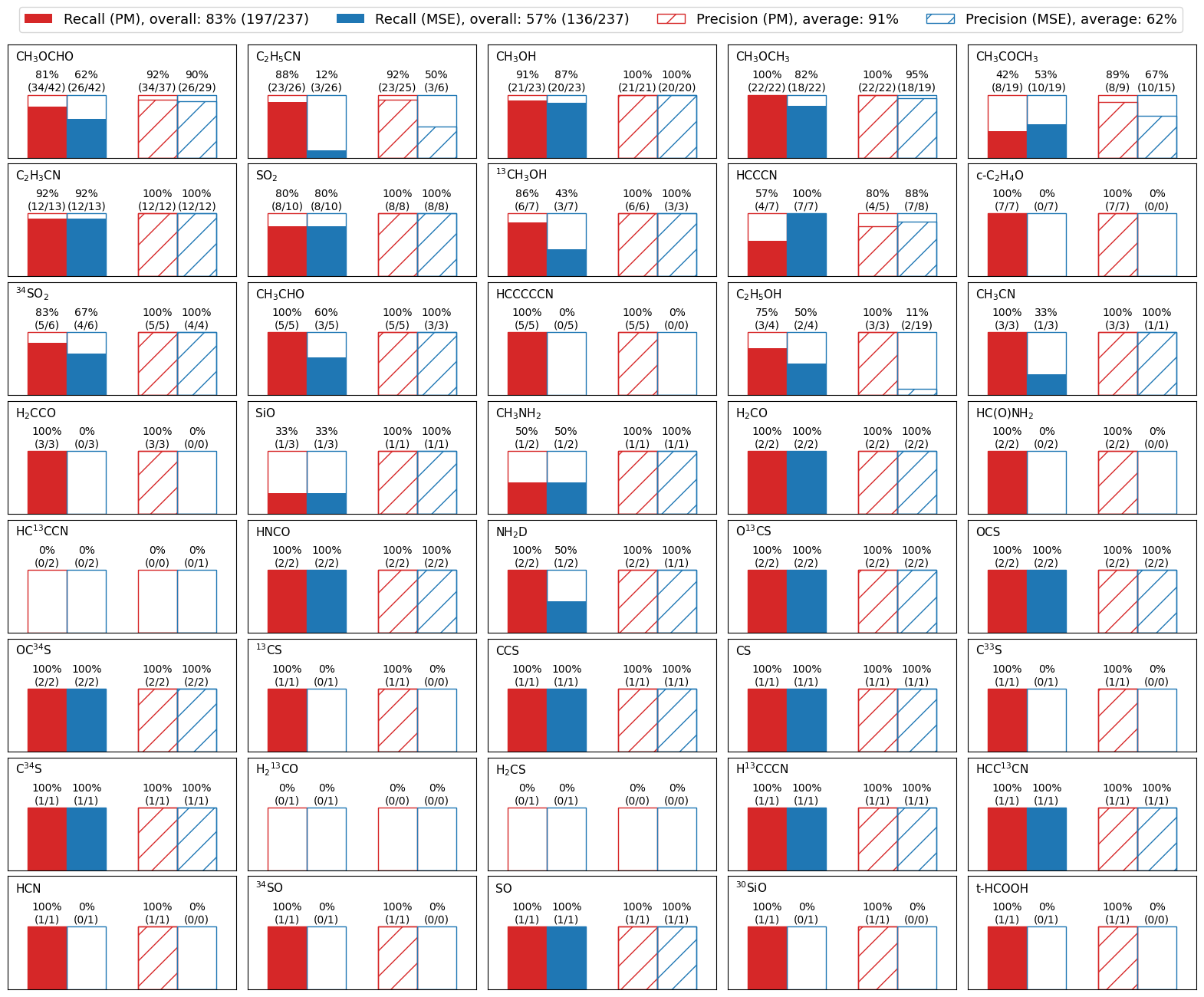}
	\centering
    \caption{Recall and precision of species identified in \textit{OrionKL-Tianma}. The red and blue bars represent the results using the peak matching and $\chi^2$ loss functions respectively. Definitions of recall and precision are provided in Section \ref{sec:metrics}.}
    \label{fig:recall_tianma}
\end{figure*}

In terms of \textit{OrionKL-Tianma}, we demonstrate the best-fitting spectra in Figure \ref{fig:spec_tianma}, the recall and precision metrics in Figure \ref{fig:recall_tianma}. The best-fitting spectrum using $\chi^2$ shows considerable wrong features, and a potential reason is that the observed spectrum consists of over 100 strong radio recombination lines, which disturbs the fitting. In contrast, the result using the peak matching loss function well agrees with the observed spectrum, indicating the robustness of our method.
\par
\begin{figure*}
	\includegraphics[width=\textwidth]{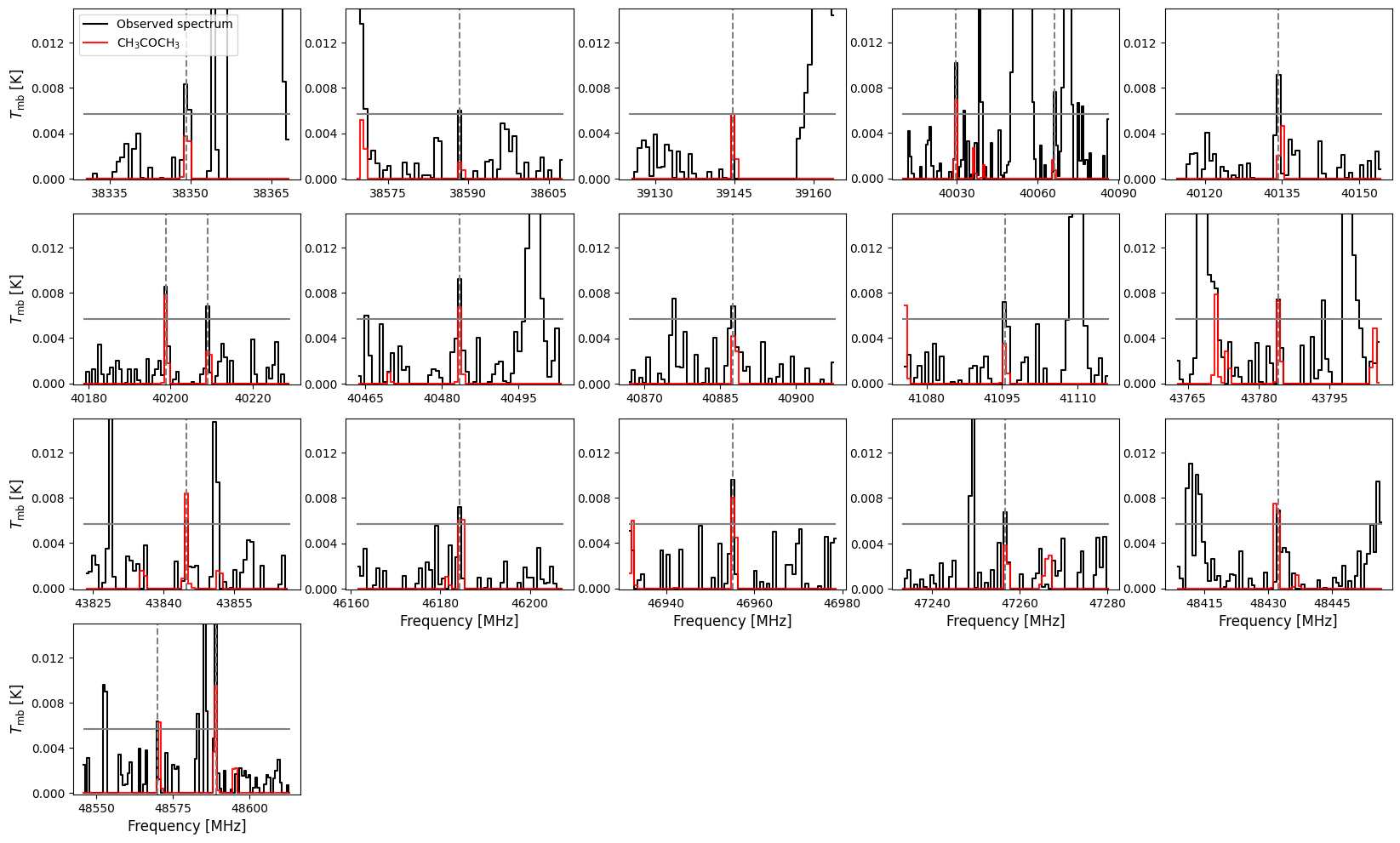}
	\centering
    \caption{Line identification of acetone (CH$_3$COCH$_3$) in \textit{OrionKL-Tianma}. The black solid lines show the observed spectrum, and the black dashed lines indicate the CH$_3$COCH$_3$ spectral lines identified by \citet{2022ApJS..263...13L}. The gray horizontal lines represent the threshold used by our algorithm to identify a peak. The red lines show the best-fitting spectrum using the peak matching loss. While the best-fitting spectrum predicts most peaks at the identified lines by \citet{2022ApJS..263...13L}, some peaks are too faint to be identified by our peak finder. This explains the low recall of CH$_3$COCH$_3$, as shown in Figure \ref{fig:recall_tianma}.}
    \label{fig:tianma_CH3COCH3}
\end{figure*}

The result using the peak matching loss function only identifies less than half acetone (CH$_3$COCH$_3$) lines, missing 11 lines. We demonstrate the issue in Figure \ref{fig:tianma_CH3COCH3}. In fact, our method reproduces the right features, while these features are slightly below the threshold of our peak finding algorithm. Consequently, these features are not counted in computing the recall. The similar problem is also found in the results of CH$_3$OCHO, C$_2$H$_5$CN, and $^{34}$SO2.
\par
In addition, as discussed by \citet{2022ApJS..263...13L}, the detected SiO emission is complex and involves masers, which cannot be described using only one velocity component. Therefore, our algorithm fails to identify all the SiO lines.
\par
\begin{figure*}
	\includegraphics[width=\textwidth]{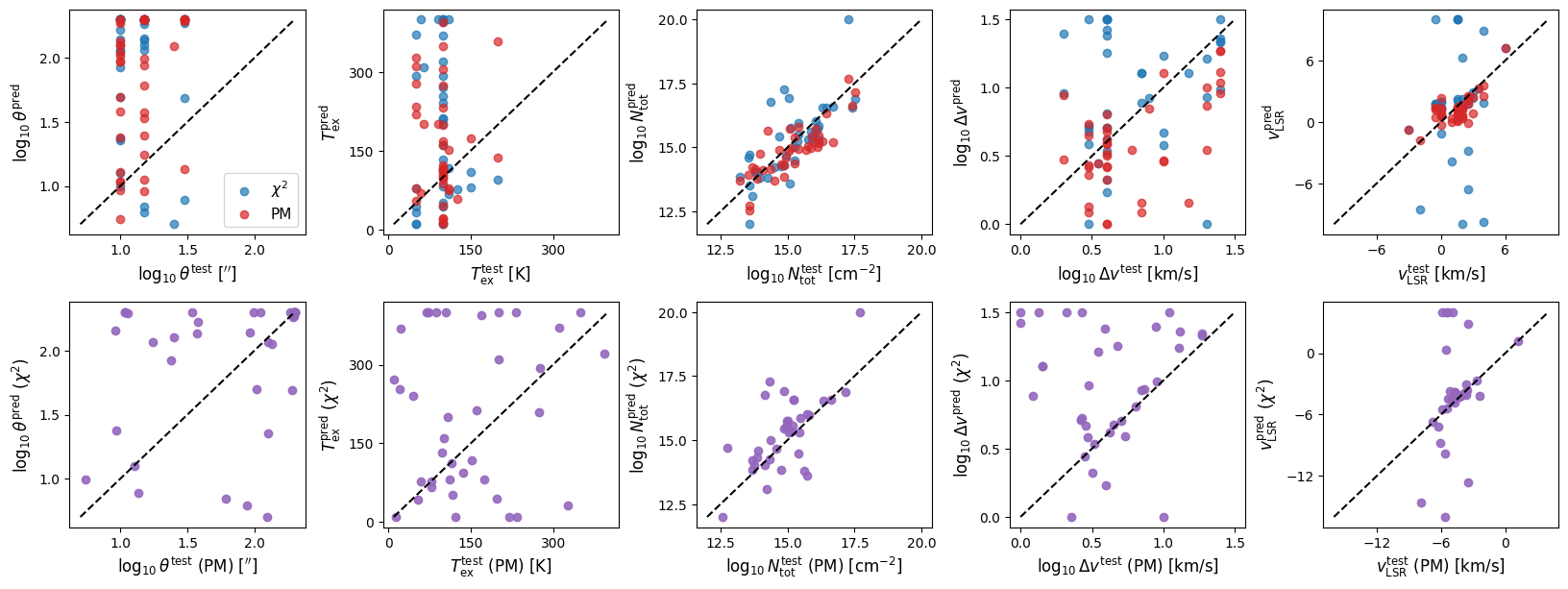}
	\centering
    \caption{Upper panels: comparison of the parameters estimated by \cite{2022ApJS..263...13L}, labeled "test", with our best-fitting parameters, labeled "pred". When the original study suggests multiple components, we compare the component with the highest integrated intensity. The red and blue dots represent the results based on the peak matching and $\chi^2$ loss functions respectively. Each data point compares the parameters of a molecule at one state. \cite{2022ApJS..263...13L} adopted fixed values for source size and excitation temperature, and fit the column density, velocity width, and velocity offset. Lower panels: comparison of the parameters derived using the peak matching and $\chi^2$ loss functions.}
    \label{fig:params_tianma}
\end{figure*}

The upper panel of Figure \ref{fig:params_tianma} compares the parameters estimated by \citet{2022ApJS..263...13L} with our best-fitting parameters. When the original study suggested multiple components of a molecule, we compare the component with the highest integrated intensity. In Figure \ref{fig:params_tianma}, for the result based on the peak matching loss, we find that the column density and velocity offset are generally consistent, whereas the source size, excitation temperature, and velocity width are inconsistent. For each molecule, \citet{2022ApJS..263...13L} fit only the column density, velocity width, and velocity offset, while used fixed values for the source size and excitation temperature, which could account for the discrepancies in the source size and excitation temperature.
\par
In the lower panel of Figure \ref{fig:params_tianma}, we compare the fitting parameters obtained with the peak matching loss and the $\chi^2$ loss functions. The predicted values for the column density, and velocity offset are correlated, while the other parameters display significant variability. In particular, the results obtained with the $\chi^2$ loss tends to predict broader velocity widths.

\subsubsection{SgrB2M-IRAM} \label{sec:b2m}
\begin{figure*}
	\includegraphics[width=\textwidth]{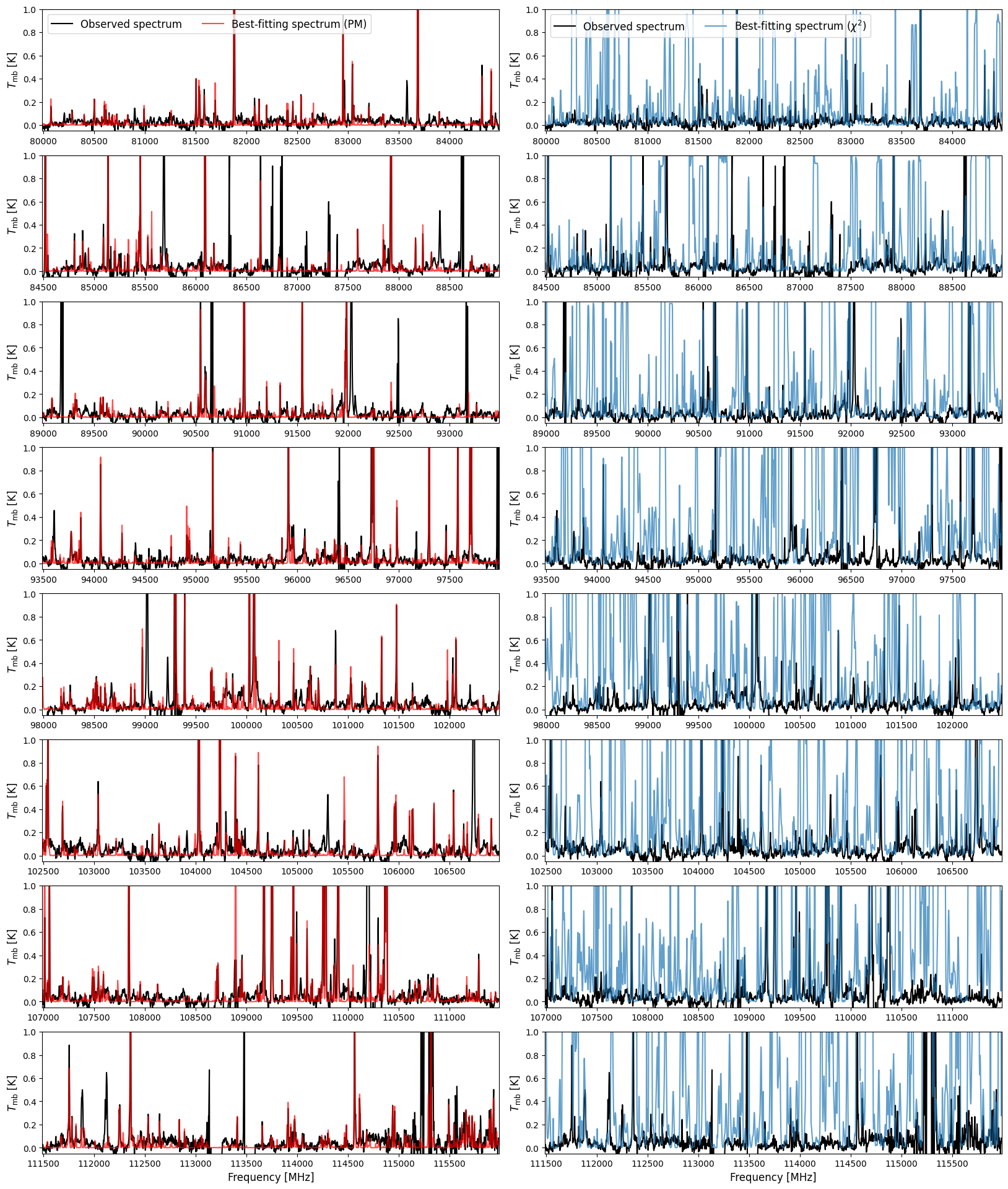}
	\centering
    \caption{Best fitting results for \textit{SgrB2M-IRAM}. Black lines represent the observed spectrum. Left and right panels show the results using the peak matching and $\chi^2$ loss functions respectively. The peak matching loss function is described in Section \ref{sec:pm_loss}.}
    \label{fig:spec_b2m}
\end{figure*}

\begin{figure*}
	\includegraphics[width=\textwidth]{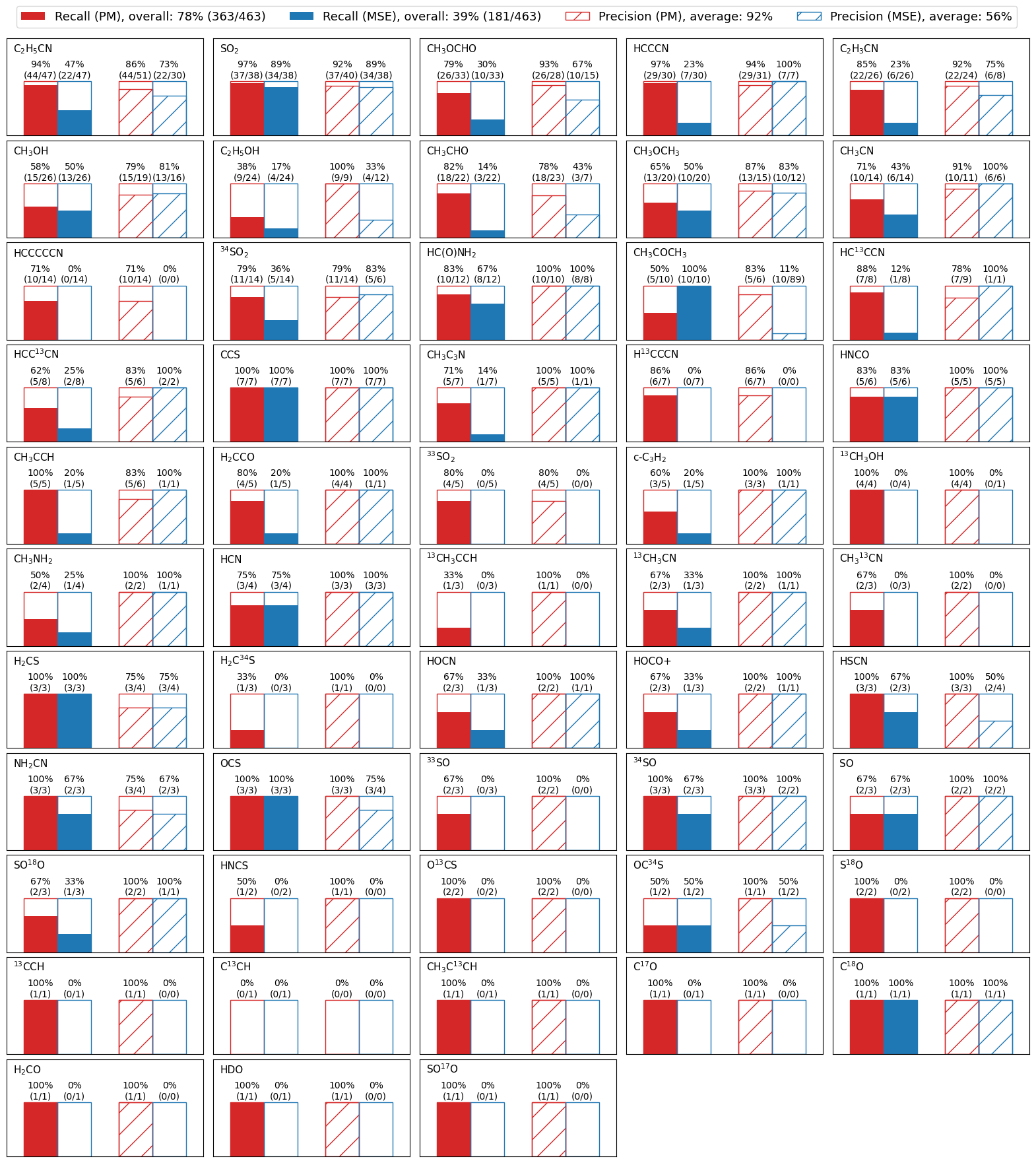}
	\centering
    \caption{Recall and precision of species identified in \textit{SgrB2M-IRAM}. The red and blue bars represent the results using the peak matching and $\chi^2$ loss functions respectively. Definitions of recall and precision are provided in Section \ref{sec:metrics}.}
    \label{fig:recall_b2m}
\end{figure*}

For \textit{SgrB2M-IRAM}, Figure \ref{fig:spec_b2m} shows the best-fitting spectra, and Figure \ref{fig:recall_b2m} illustrates the recall and precision values for each molecule. For the results based on the $\chi^2$ loss function, the best-fitting spectrum is fundamentally flawed, with a overall recall of 39\% and an average precision of 56\%. In contrast, for this line rich region, our result using the peak matching loss fits the observed spectrum well, achieving an overall recall of 78\% and an average precision of 92\%.
\par
\begin{figure*}
	\includegraphics[width=\textwidth]{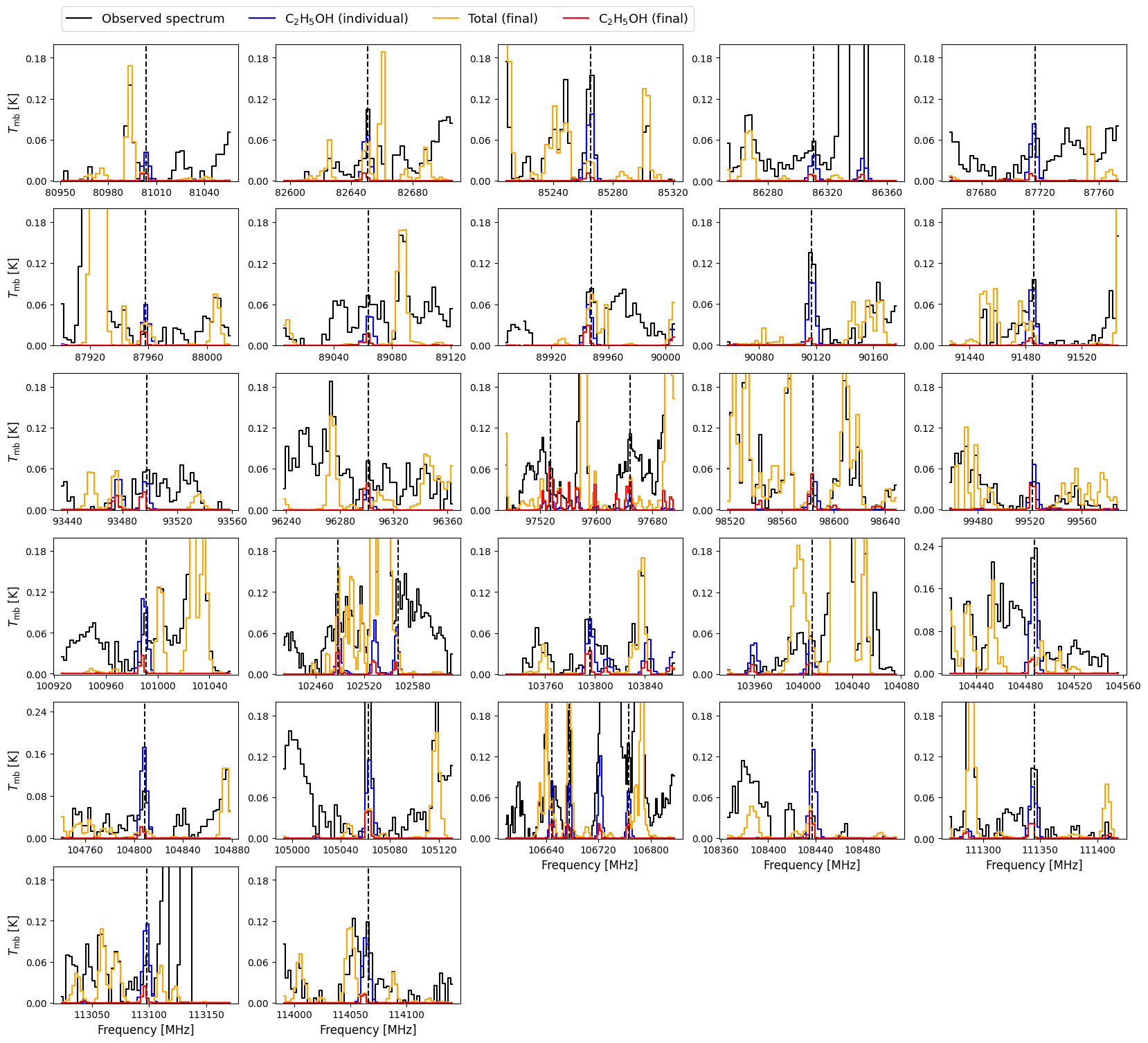}
	\centering
    \caption{Line blending analysis of ethanol (C$_2$H$_5$OH) in \textit{SgrB2M-IRAM}. The black solid lines represent the observed spectrum, and the black dashed lines indicate the C$_2$H$_5$OH spectral lines identified by \citet{2013A&A...559A..47B}. The blue lines show the best-fitting spectrum of C$_2$H$_5$OH obtained in the individual fitting phase. The red solid lines illustrate the spectrum of C$_2$H$_5$OH as a part of the combined spectrum. The final combined spectrum is shown as the yellow lines. While the final result show a low recall of C$_2$H$_5$OH, most lines can be recovered in the individual fitting phase, implying line blending.}
    \label{fig:b2m_C2H5OH}
\end{figure*}

Our results show a low recall of ethanol (C$_2$H$_5$OH) at 38\%, potentially due to line blending. To demonstrate the problem, we present the best-fitting results using the peak matching loss in Figure \ref{fig:b2m_C2H5OH}. The blue lines represent the best-fitting spectrum of C$_2$H$_5$OH from the individual fitting phase, aligning with most of the C$_2$H$_5$OH lines (indicated by black vertical lines) as identified by \cite{2013A&A...559A..47B}. However, these lines are absent in the final combined spectrum, implying that the fitting spectrum of C$_2$H$_5$OH is influenced by contributions from other species.
\par
\begin{figure*}
	\includegraphics[width=\textwidth]{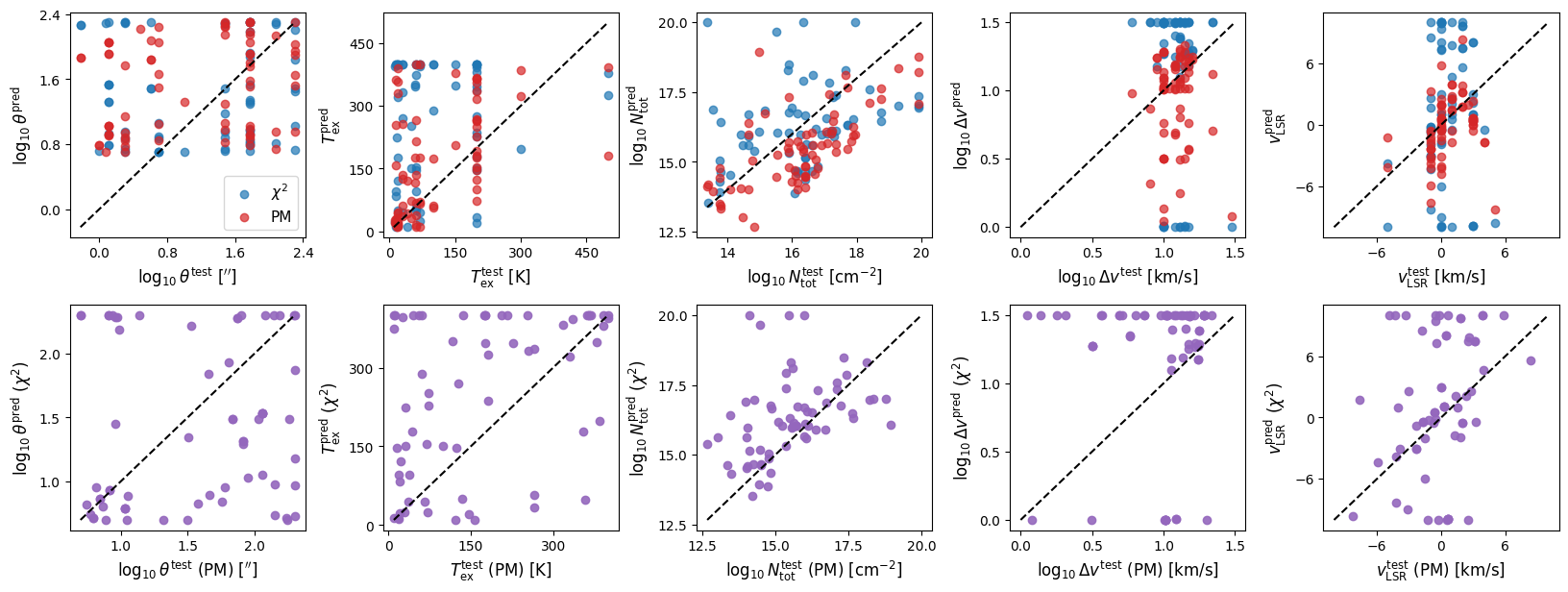}
	\centering
    \caption{Comparison of the parameters estimated by \cite{2013A&A...559A..47B}, labeled "test", with the best-fitting parameters obtained in this work, labeled "pred". Whenever the original study suggests multiple components, the component with the highest column density is compared. Each data point compares the parameters of a molecule at one state. The red and blue dots represent the results based on the peak matching and $\chi^2$ loss functions respectively.}
    \label{fig:params_b2m}
\end{figure*}

In Figure \ref{fig:params_b2m}, we compare the estimated parameters from \citet{2013A&A...559A..47B} with our results. When the original study suggested multiple components of a molecule, we compare the component with the highest integrated intensity.\citet{2013A&A...559A..47B} also generated the spectra using \textsc{xclass}, and their parameters were adjusted through visual comparison with the observed spectrum. While a correlation is found for the column density, our results based on the peak matching loss tend to underestimate it by 1 - 2 dex. A potential reason could be that our algorithm tends to adopt larger source sizes than \citet{2013A&A...559A..47B}. Moreover, no significant correlations can be found for the other parameters. In particular, the predicted $\delta v_\text{LSR}$ values exhibit large uncertainties. We estimated that the standard deviation of the velocity offset errors, $v^\text{test}_\text{LSR} - v^\text{pred}_\text{LSR}$, is $\sim$ 3.0 km/s, larger than the standard deviation of $v^\text{test}_\text{LSR}$, at 1.0 km/s. This could explain the absence of a correlation between $v^\text{test}_\text{LSR}$ and $v^\text{pred}_\text{LSR}$. However, we consider the errors in the predicted velocity offsets to be acceptable, since the mean velocity width of the test samples is 12.4 km/s, which is four times the standard deviation of the velocity offset errors. Additionally, we verified that the significant deviations of the predicted $\delta v_\text{LSR}$ result from poor fitting. For instance, as previously discussed above, the metrics for C$_2$H$_5$OH are unsatisfactory. The predicted $\delta v_\text{LSR}$ for C$_2$H$_5$OH is 5.8 km/s, whereas \citet{2013A&A...559A..47B} reported a value of 0.0 km/s.

\subsubsection{SgrB2N-IRAM}
\begin{figure*}
	\includegraphics[width=\textwidth]{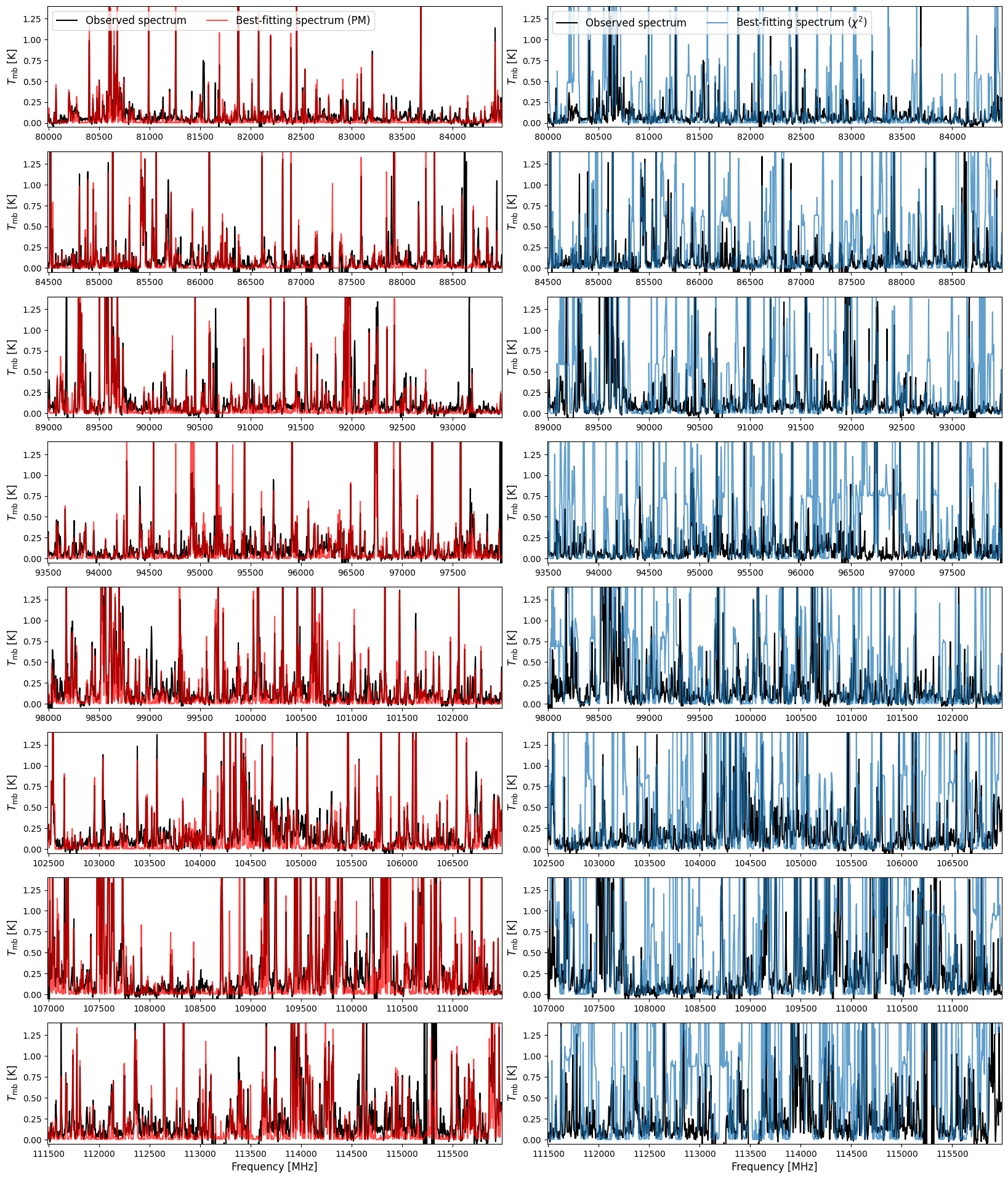}
	\centering
    \caption{Best fitting results for \textit{SgrB2N-IRAM}. Black lines represent the observed spectrum. Left and right panels show the results using the peak matching and $\chi^2$ loss functions respectively. The peak matching loss function is described in Section \ref{sec:pm_loss}.}
    \label{fig:spec_b2n}
\end{figure*}

\begin{figure*}
	\includegraphics[width=.9\textwidth]{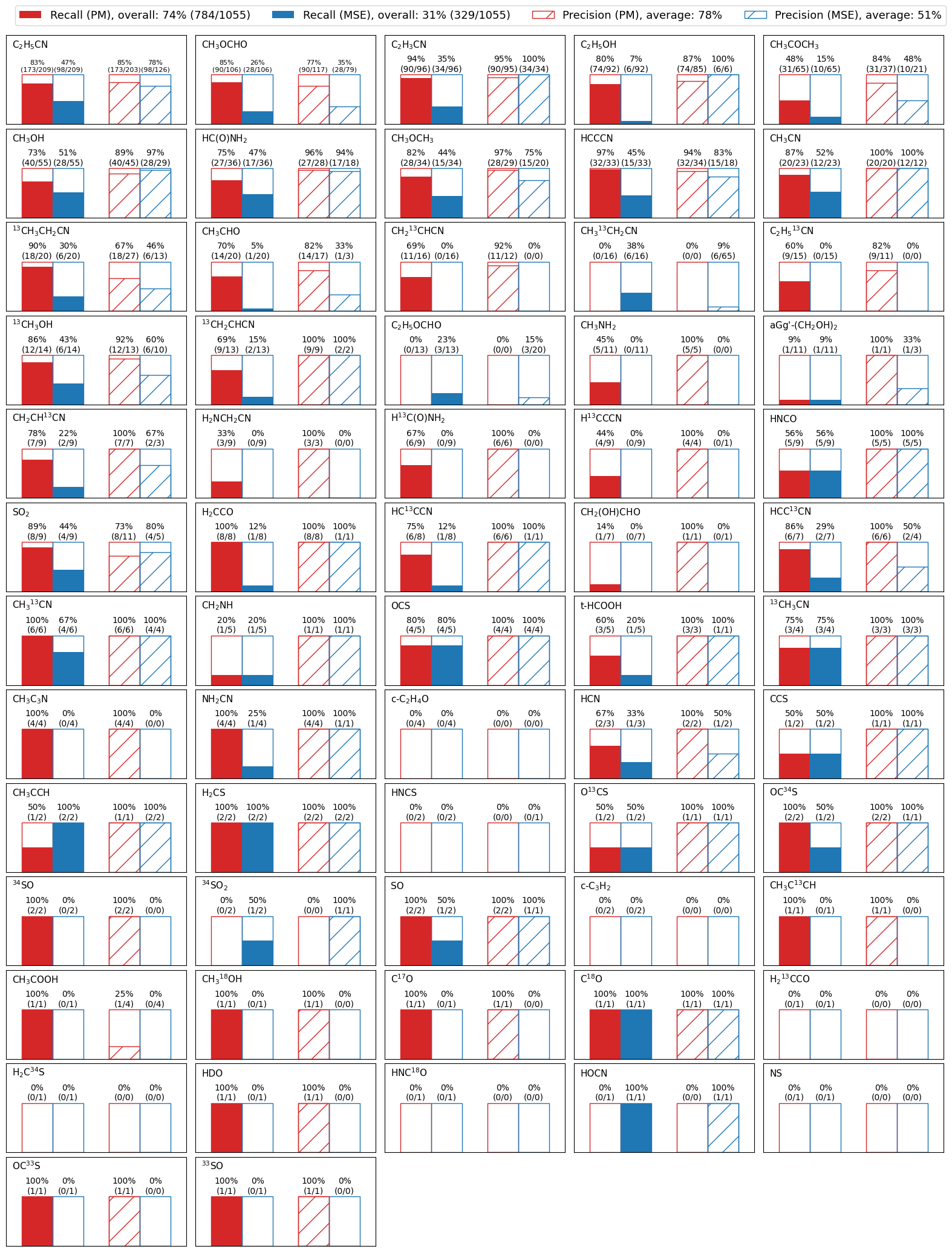}
	\centering
    \caption{Recall and precision of species identified in \textit{SgrB2N-IRAM}. The red and blue bars represent the results using the peak matching and $\chi^2$ loss functions respectively. Definitions of recall and precision are provided in Section \ref{sec:metrics}.}
    \label{fig:recall_b2n}
\end{figure*}

\textit{SrgB2N-IRAM} is the most challenging test data in this study, characterized by a line density of 59.2 GHz$^{-1}$. In Figure \ref{fig:spec_b2n}, the best-fitting spectrum based on the peak matching loss is in good agreement with the observed spectrum, while the outcome derived from the $\chi^2$ loss is inconclusive. Regarding the metrics depicted in Figure \ref{fig:recall_b2n}, the peak matching loss function demonstrates superior performance compared to the $\chi^2$ loss function, with the overall recall rising from 31\% to 74\%, and the average precision from 51\% to 78\%.
\par
\begin{figure*}
	\includegraphics[width=\textwidth]{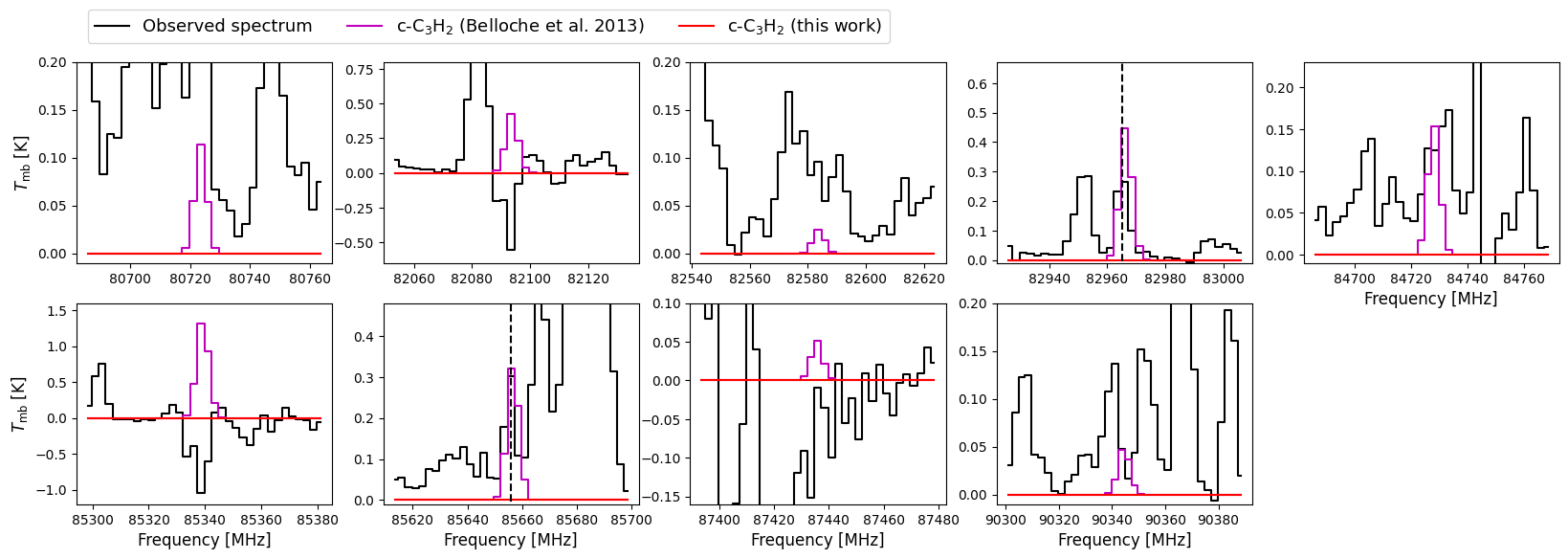}
	\centering
    \caption{Line identification of cyclopropenylidene (c-C$_3$H$_2$) by \citet{2013A&A...559A..47B}. The black solid lines represent the observed spectrum, and the black dashed lines indicate the spectral lines of c-C$_3$H$_2$ identified by \citet{2013A&A...559A..47B}. The purple lines represent the spectrum generated using the parameters suggested by \citet{2013A&A...559A..47B}. Each panel displays a peak from the model spectrum. These plots imply that absorption should be taken into account when fitting the c-C$_3$H$_2$ spectrum.}
    \label{fig:b2n_C3H2}
\end{figure*}

The identification of cyclopropenylidene (c-C$_3$H$_2$) is a special case, affected by absorption in the observed spectrum. Figure \ref{fig:b2n_C3H2} displays the c-C$_3$H$_2$ spectrum generated using the parameters suggested by \cite{2013A&A...559A..47B}. Absorption features are found at the locations of three peaks, at 82094 MHz, 85339 MHz, and 87436 MHz. Proper modeling of the absorption component would lead to the attenuation of these three peaks. However, in this study, absorption is not taken into account, and the affected spectrum is set to zero. As a result, these three peaks are classified as false positives by our algorithm, leading to an increased loss. Consequently, the identification c-C$_3$H$_2$ becomes challenging for our algorithm.
\par
In addition, our algorithm fails to identify HNCS and HNC$^{18}$O. We find it difficult to reproduce the spectral lines of these molecules using the parameters suggested by \citet{2013A&A...559A..47B}. A potential reason could be that different spectral data were used.
\par
\begin{figure*}
	\includegraphics[width=\textwidth]{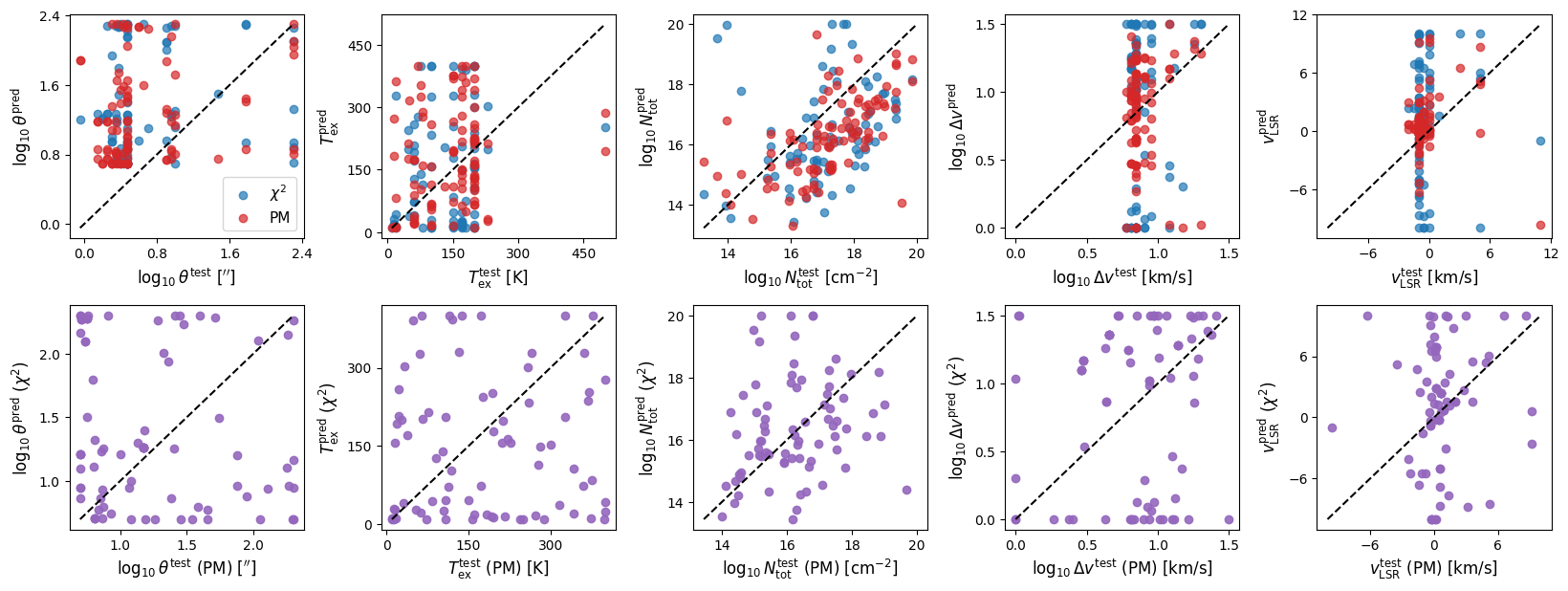}
	\centering
    \caption{Comparison of the parameters estimated by \cite{2013A&A...559A..47B}, labeled "test", with the best-fitting parameters obtained in this work, labeled "pred". Whenever the original study suggests multiple components, the component with the highest column density is compared. Each data point compares the parameters of a molecule at one state. Red and blue dots represent the results based on the peak matching and $\chi^2$ loss functions respectively.}
    \label{fig:params_b2n}
\end{figure*}

In Figure \ref{fig:params_b2n}, we compare the parameters suggested by \citet{2013A&A...559A..47B} with our best-fitting parameters. Both results employ \textsc{xclass} to generate model spectra. Similar to \textit{SrgB2M-IRAM}, we can only find correlation for the column density, and the other parameters show discrepancies. The potential reasons have been discussed in Section \ref{sec:b2m}.

\subsection{Computational cost}
\begin{figure}
	\includegraphics[width=.9\columnwidth]{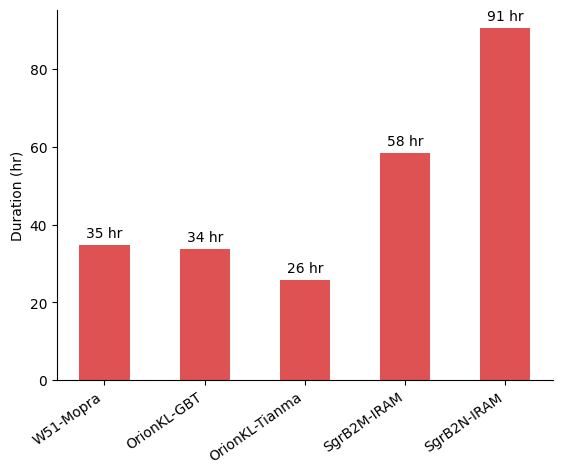}
	\centering
    \caption{Computation time for our automated line identification algorithm using the peak matching loss function.}
    \label{fig:time}
\end{figure}

Figure \ref{fig:time} demonstrates the computation time of our automated line identification method with the peak matching loss function. For all line surveys analyzed in this study, our algorithm is executed using 28 2.6 GHz CPUs in parallel. The computational cost varies from 26 to 91 hours, depending on the complexity of the observed spectrum. The computation resources required by our method are relatively demanding. We find that the majority of the computational time is devoted to calling \textsc{xclass}, while the time allocated to optimization algorithm itself and the calculation of the peak matching loss function is minimal. Accordingly, enhancing the efficiency of the spectral line model could expedite our line identification pipeline.

\section{Summary} \label{sec:sum}
To summarize, we develop a framework for automated line identification of interstellar molecules. Our main contributions are summarized as follows.
\begin{itemize}
    \item We have shown that the widely used $\chi^2$ loss function tends to produce unsatisfactory results under minimal human supervision. To address this issue, we propose the peak matching loss function, which significantly enhances the robustness of spectral fitting. When applying to our automated line identification framework, the results obtained with the peak matching loss function are consistently better than those from the $\chi^2$ loss function. It is worth noting that the peak matching loss function is an independent component in our method, and in principle can be incorporated into other codes to improve the fitting results.
    \item We develop a greedy algorithm to combine the fitting results of individual molecules. The algorithm is efficient and interpretable. 
\end{itemize}
\par
We evaluate our algorithm using published data of five line surveys, including line rich regions in Sgr B2. Our algorithm achieves an overall recall of 74\% - 93\% and an average precision of 78\% - 92\%. The errors of our algorithm are mainly due to the following factors:
\begin{itemize}
    \item Due to the stochastic nature of particle swarm optimization, our results involve random errors.
    \item While our method assumes a single physical component for all molecules for simplicity, certain molecules, e.g. methanol, might be better described by incorporating multiple components.
    \item Line blending can result in ambiguity and multiple possible solutions.
    \item Absorption in the spectrum can affect spectral fitting.
\end{itemize}
\par
We also compare our best-fitting parameters with the previous studies. Whereas our best-fitting spectra are in good agreement with the observations, the estimated parameters are in general inconsistent with the previous studies. We find a correlation only for the column density, with an error of two orders of magnitude. We attribute this discrepancy to the inherent degeneracy between the source size, excitation temperature, and column density. We suggest that the uncertainties of the fitting parameters can be better studied using mock data where the true parameters are known. Such studies could constitute future work.
\par
Our method contributes to significantly reducing the workload of spectral line identifications. The results produced by our method provide an excellent starting point for a more detailed analysis.

\section*{acknowledgments}
This work was supported by the National Natural Science Foundation of China (Grant No. 12373026), the Leading Innovation and Entrepreneurship Team of Zhejiang Province of China (Grant No. 2023R01008), the Key R\&D Program of Zhejiang, China (Grant No. 2024SSYS0012), and Young Scientists Fund of the National Natural Science Foundation of China (Grant No.  12403030). The computation is performed at the Shuguang supercomputer in Zhejiang Lab. We thank the reviewer for providing a constructive report, which improved the quality of the paper.
%\end{acknowledgments}


\vspace{5mm}
\facilities{GBT 100 m, IRAM 30 m, Mopra 22 m, Tianma 65 m (TMRT).}


\software{\textsc{matplotlib} \citep{Hunter:2007}, \textsc{numpy} \citep{harris2020array}, \textsc{pandas} \citep{mckinney-proc-scipy-2010,reback2020pandas}, \textsc{scipy} \citep{2020SciPy-NMeth}, \textsc{xclass} \citep{2017A&A...598A...7M}.
}



%\appendix

%\section{Online Tables}


\bibliography{references}{}
\bibliographystyle{aasjournal}



\end{document}